\documentclass[letterpaper,english,letterpaper,twocolumn,floatfix,groupedaddress,superscriptaddress]{revtex4}
\usepackage{mathptmx}
\usepackage[T1]{fontenc}
\usepackage[latin9]{inputenc}
\usepackage{amsmath}
\usepackage{amssymb}
\usepackage{mathrsfs}
\usepackage{graphicx}
\usepackage{subfigure}



\usepackage{babel}

\begin{document}

\title{Quantum Computing Without Wavefunctions: Time-Dependent Density Functional Theory for Universal Quantum Computation}

\author{David G. Tempel}

\address{Department of Physics, Harvard University,
17 Oxford Street, 02138, Cambridge, MA}

\author{Al\'an Aspuru-Guzik}

\address{Department of Chemistry and Chemical Biology, Harvard University,
12 Oxford Street, 02138, Cambridge, MA}

\maketitle



\textbf{By balancing both accuracy and efficiency, density functional theory (DFT)~\cite{HK_theorem, Kohn_Sham} and its time-dependent extension (TDDFT)~\cite{runge gross, TDDFT_review} have become arguably the most employed methods in computational physics and chemistry. DFT and TDDFT are based on rigorous theorems, which reformulate many-electron quantum mechanics using the simple one-electron density as the basic variable of interest rather than the complicated many-electron wavefunction. In this letter we prove that the theorems of TDDFT can be applied to a class of qubit Hamiltonians that are universal for quantum computation. In a similar spirit to DFT and TDDFT for electronic Hamiltonians, the theorems of TDDFT applied to universal Hamiltonians allow us to think of single-qubit expectation values as the basic variables in quantum computation and information theory, rather than the wavefunction. From a practical standpoint this also opens the possibility of approximating observables of interest in quantum computations directly in terms of single-qubit quantities (i.e. as density functionals). Additionally, we demonstrate that TDDFT provides an exact prescription for simulating universal Hamiltonians with other universal Hamiltonians that have different, and possibly easier-to-realize two-qubit interactions.}

We begin by briefly reviewing TDDFT for a system of N-electrons described by the Hamiltonian
\begin{equation}
\hat{H}(t) = \sum_{i=1}^N \frac{\hat{p}^2_i}{2m} + \sum_{i<j}^N w(|\hat{\mathbf{r}}_i - \hat{\mathbf{r}}_j|) + \int  v(\mathbf{r}, t) \hat{n}(\mathbf{r}) d^3 \mathbf{r},
\label{electron_Hamiltonian}
\end{equation}
where $\hat{\mathbf{p}}_i$ and $\hat{\mathbf{r}}_i$ are respectively the position and momentum  operators of the ith electron, $w(|\hat{\mathbf{r}}_i - \hat{\mathbf{r}}_j|)$ is the electron-electron repulsion and $v(\mathbf{r}, t)$ is a time-dependent one-body scalar potential which includes the potential due to nuclear charges as well as any external fields. $\hat{n}(\mathbf{r}) = \sum_i^N \delta(\mathbf{r} - \hat{\mathbf{r}}_i)$ is the electron density operator, whose expectation value yields the one-electron probability density. The first basic theorem of TDDFT, known as the "Runge-Gross (RG) theorem"~\cite{runge gross}, establishes a one-to-one mapping between the expectation value of $\hat{n}(\mathbf{r})$ and the scalar potential $v(\mathbf{r}, t)$ and therefore through the time-dependent Schr{\"o}dinger equation, a one-to-one mapping between the density and the wavefunction. The RG theorem implies the remarkable fact that in principle, the one-electron density contains the same information as the many-electron wavefunction. The second basic TDDFT theorem known as the "van Leeuwen (VL) theorem"~\cite{van leeuwen} gives a prescription for constructing an auxiliary system with a different and possibly simpler electron-electron repulsion $w'(|\hat{\mathbf{r}}_i - \hat{\mathbf{r}}_j|)$, which simulates the density evolution of the original Hamiltonian in Eq.~\ref{electron_Hamiltonian}. When $w'(|\hat{\mathbf{r}}_i - \hat{\mathbf{r}}_j|)=0$, this auxiliary system is referred to as the "Kohn-Sham system"~\cite{Kohn_Sham} and due to it's simplicity and accuracy, is in practice used in most DFT and TDDFT calculations .

It is not obvious that the RG and VL theorems extend to qubits, which are {\sl distinguishable} spin 1/2 particles. We now prove analogous RG and VL theorems for a system of N qubits described by the universal 2-local Hamiltonian~\cite{Bose_Benjamin, Bose_Benjamin_PRA},
\begin{equation}
\hat{H}(t) = \sum_{i=1}^{N-1} J^{\perp}_{i, i+1} (\hat{\sigma}_i^{x} \hat{\sigma}_{i+1}^{x} + \hat{\sigma}_i^{y} \hat{\sigma}_{i+1}^{y}) + \sum_{i =1}^{N-1} J^{\parallel}_{i, i+1} \hat{\sigma}_i^{z} \hat{\sigma}_{i+1}^{z} + \sum_{i=1}^N h_i(t) \hat{\sigma}_i^z.
\label{qubit Hamiltonian}
\end{equation}
Here, $\hat{\sigma}_i^{x}$, $\hat{\sigma}_i^{y}$,  $\hat{\sigma}_i^{z}$ are Pauli operators for the ith qubit, $h_i(t)$ are local applied fields arbitrarily chosen along the z-axis and $J^{\parallel}_{i, i+1}$ and $J^{\perp}_{i, i+1}$ are two-qubit interaction terms respectively parallel and perpendicular to the direction of the fields. The above Hamiltonian describes an open chain of N qubits arranged in a one-dimensional array, with each qubit interacting with its nearest neighbors. In Refs.~\cite{Bose_Benjamin, Bose_Benjamin_PRA}, it was shown that by appropriately tuning the local fields in Eq.~\ref{qubit Hamiltonian}, one can realize any two qubit gate, which in turn can be employed to perform universal quantum computation when combined with singe-qubit rotations. Although Eq.~\ref{qubit Hamiltonian} describes quantum computing with a fixed two-qubit interaction, by applying appropriate local fields we can generate effective Hamiltonians describing systems with tunable two-qubit interactions as well. This can be viewed as a time-dependent version of the widely employed method of Gadgets~\cite{Gadget_paper}. In Eq.~\ref{qubit Hamiltonian}, the case where $J^{\perp}_{i, i+1} = J^{\parallel}_{i, i+1}$ yields the Heisenberg Hamiltonian which describes exchange coupled spins in solid state arrays or quantum dots in heterostructures~\cite{Nature_paper}. The situation $J^{\perp}_{i, i+1} \neq J^{\parallel}_{i, i+1}$ yields the XXZ Hamiltonian, used to model electronic qubits on liquid Helium~\cite{Science_paper} or solid-state systems with anisotropy due to spin-orbit coupling~\cite{Lidar_PRL}, while the limit $J^{\parallel}_{i, i+1} = 0$ yields the XY model describing superconducting Josephson junction qubits~\cite{Nature_paper_2}. 


\begin{figure}
\begin{centering}
\includegraphics[width=1.0\columnwidth]{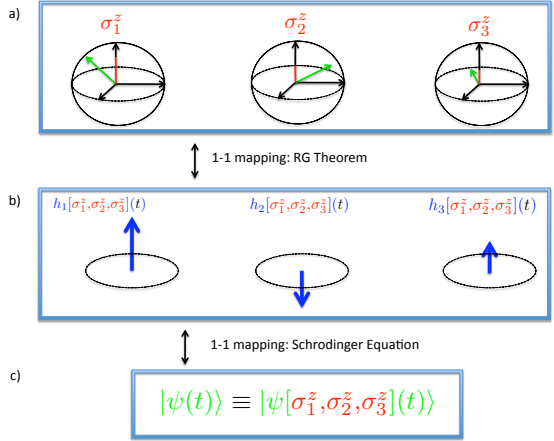} 
\end{centering}
\caption{\textbf{Runge-Gross theorem for a 3 qubit example} - The set of expectation values $\{ \sigma_1^z , \sigma_2^z , ... \sigma_N^z  \}$, defined by the the Bloch vector components of each qubit along the  z-axis in (a), is uniquely mapped onto the set of local fields $\{ h_1, h_2, ... h_N \}$ in (b) through the RG theorem. Then, through the Schr{\"o}dinger equation, the set of fields is uniquely mapped onto the wavefunction. These two mappings together imply that the N-qubit wavefunction in (c) is in fact a unique functional of the set of expectation values $\{ \sigma_1^z , \sigma_2^z , ... \sigma_N^z  \}$.}
\label{RG_theorem}
\end{figure}

We now state the equivalent RG theorem for quantum computation with the Hamiltonian in Eq.~\ref{qubit Hamiltonian}: 
\vspace{.3cm}
\\ \textit{Theorem - For a given initial state $|\psi(0)\rangle$ evolving to $|\psi(t)\rangle$ under the Hamiltonian in Eq.~\ref{qubit Hamiltonian} and with $J^{\parallel}_{i, i+1}$ and $J^{\perp}_{i, i+1}$ fixed, there exists a one-to-one mapping between the set of expectation values $\{ \sigma_1^z , \sigma_2^z , ... \sigma_N^z  \}$ and the set of local fields $\{ h_1, h_2, ... h_N \}$ over a given interval [0 ,t].} \vspace{.3cm} \\ Here, we have defined $\sigma_i^z  \equiv \langle \psi(t)| \hat{\sigma}_i^z | \psi(t) \rangle$ as the expectation value of the component of the ith qubit along the field direction (z-axis). A detailed proof together with a more rigorous discussion of the conditions on the theorem are provided in the supplementary material. The RG theorem implies that the set of local fields can be written as unique functionals of the set of expectation values $\{ \sigma_1^z , \sigma_2^z , ... \sigma_N^z  \}$, as illustrated in the first part of Figure~\ref{RG_theorem}. Since the solution to the time-dependent Schr{\"o}dinger equation is unique and $J^{\perp}_{i, i+1}$ and $J^{\parallel}_{i, i+1}$ are fixed, the wavefunction is a unique functional of the local fields. i.e. $|\psi(t)\rangle \equiv |\psi[h_1, h_2, ... h_N](t) \rangle$, where the square brackets denote that $\psi$ is a functional of the set $\{ h_1, h_2, ... h_N \}$ over the interval [0,t]. This fact, combined with the RG theorem allows us to state a corollary, which is the first central result of this letter:
\vspace{.3cm}
\\ \textit{Corollary - There exists a one-to-one mapping between the set of expectation values $\{ \sigma_1^z , \sigma_2^z , ... \sigma_N^z  \}$ and the N-qubit wavefunction $|\psi(t)\rangle$ on the interval [0 ,t].} \vspace{.3cm} \\ The above corollary implies the counterintuitive fact that the full N-qubit wavefunction, which lives in a $2^N$ dimensional Hilbert space, is a unique functional of only the N components of each qubit along the z-axis. i.e. 
\begin{equation}
|\psi(t)\rangle \equiv |\psi[\sigma_1^z , \sigma_2^z, ... \sigma_N^z] (t) \rangle.
\label{wavefunction}
\end{equation}

Although the RG theorem does not tell us an explicit functional form for $\psi$, it has profound conceptual implications from a quantum information perspective. At first glance, it might appear that the set $\{ \sigma_1^z , \sigma_2^z , ... \sigma_N^z  \}$ contains much less information than the full wavefunction, since projective measurements needed to obtain $\{ \sigma_1^z , \sigma_2^z , ... \sigma_N^z  \}$ would typically imply that information about non-commuting obesrvables is lost. However, since the wavefunction completely specifies all properties of the system, Eq.~\ref{wavefunction} implies that even properties depending on non-commuting observables such as entanglement and phase information are in fact unique functionals of the set of expectation values $\{ \sigma_1^z , \sigma_2^z , ... \sigma_N^z  \}$. 


From a practical standpoint, the RG theorem implies that all observables can \textit{directly} be approximated as functionals of single-qubit expectation values, without regard for the wavefunction. Although the set of expectation values $\{ \sigma_1^z , \sigma_2^z , ... \sigma_N^z  \}$ in principle contains all of the quantum information in $\psi$, extracting this information directly is not always straightforward, although in some cases it is. As an example, consider a computation involving only one flipped qubit relative to the other $N-1$ qubits having an opposite orientation. An explicit entanglement functional (as measured by concurrence~\cite{Wooters}) between any two qubits labeled $k$ and $l$ can be written very simply as \begin{equation}
E_{kl}[\sigma_1^z , \sigma_2^z, ... \sigma_N^z] (t) = \frac{1}{N-2} \prod_{m = k,l} \left[(N-3) \sigma_m^z  + \sum_{i \neq m}^N \sigma_i^z\right]^{\frac{1}{2}}.
\label{entanglement}
\end{equation}  Interestingly, this particular entanglement functional is time-local, since it depends only on the set $\{ \sigma_1^z , \sigma_2^z , ... \sigma_N^z  \}$ at a given instant in time and so we may write $E_{kl}[\sigma_1^z , \sigma_2^z, ... \sigma_N^z] (t) = E_{kl}[\sigma_1^z(t) , \sigma_2^z(t), ... \sigma_N^z(t)]$. In the more general case, observables may be non-local in time and depend on the set $\{ \sigma_1^z , \sigma_2^z , ... \sigma_N^z  \}$ over an entire interval [0,t]. Although the functional in Eq.~\ref{entanglement} is time-local, it is spatially non-local, since the entanglement between qubits k and l depends on the components of all of the other $N-2$ qubits. If one considers two flipped qubits instead of one, the entanglement functional becomes complicated and non-local in both space and time due to dependence on phases in the wavefunction (see supplemental material). Understanding the spatial and temporal non-locality of density functionals in electronic structure theory is a very active research topic~\cite{non-locality, memory}, and naturally arises here in TDDFT for quantum computation as well.



\begin{figure}
\begin{centering}
\includegraphics[width=1.0\columnwidth]{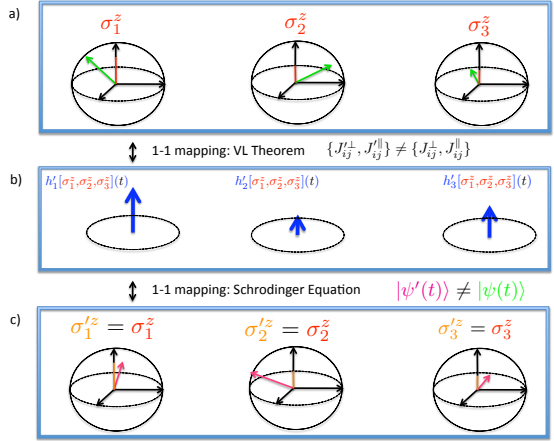} 
\end{centering}
\caption{\textbf{Van Leeuwen theorem for a 3 qubit example} - The set $\{ \sigma_1^z , \sigma_2^z , ... \sigma_N^z  \}$ (a) obtained from evolution under Eq.~\ref{qubit Hamiltonian}, is uniquely mapped to a new set of fields $\{ h'_1, h'_2, ... h'_N \}$ (b) for a Hamiltonian with different two-qubit interactions. Evolution under this new Hamiltonian returns the same expectation values $\{ \sigma_1^z , \sigma_2^z , ... \sigma_N^z  \}$, although the wavefunction is different and hence projections of the Bloch vectors along other axes are in general different (c).}
\label{VL_theorem}
\end{figure}

We now turn to the second fundamental theorem of TDDFT for universal computation, a VL-like theorem for qubits: \vspace{.3cm} \\ \textit{Theorem - Consider a given set of spin components $\{ \sigma_1^z , \sigma_2^z , ... \sigma_N^z  \}$ obtained from the wavefunction $|\psi(t)\rangle$ evolved under the Hamiltonian in Eq.~\ref{qubit Hamiltonian}.  There exists (see supplementary material for certain conditions) a Hamiltonian with different two-qubit interactions denoted $J'^{\perp}_{i, i+1}$ and $J'^{\parallel}_{i, i+1}$ and different local fields $\{ h'_1, h'_2, ... h'_N \}$, which evolves a possibly different initial state $|\psi'(0)\rangle$ to a different final state  $|\psi'(t)\rangle$ such that the condition $\{ \sigma_1'^z , \sigma_2'^z , ... \sigma_N'^z  \}$ = $\{ \sigma_1^z , \sigma_2^z , ... \sigma_N^z  \}$ is satisfied on the interval [0,t].} \vspace{.3cm} \\ Here, we have defined  $\sigma_i'^z  \equiv \langle \psi'(t)| \hat{\sigma}_i^z | \psi'(t) \rangle$.  The VL theorem allows us to obtain the set $\{ \sigma_1^z , \sigma_2^z , ... \sigma_N^z  \}$ by simulating the evolution with an auxiliary Hamiltonian having different two-qubit interactions and hence a different (and possibly simpler) wavefunction evolution as illustrated in Figure~\ref{VL_theorem}. This opens the possibility of simplifying computations by constructing simple approximations to the auxiliary fields as functionals of single-qubit expectation values, in the same sense that the exchange-correlation potential of electronic DFT and TDDFT is approximated as a functional of the one-body density in the Kohn-Sham scheme. 

\begin{figure*}[h]
\begin{centering}
$\begin{array}{c@{\hspace{1in}}c}
 \\[-0.53cm]  \text{\underline{\Large{Heisenberg interaction}} \hspace{5.0cm} \underline{\Large{XY interaction}}} \\ [0.3cm]\\ \hspace{-7.5cm} \text{a)} \hspace{8.2cm} \text{c)} \\ [-.05cm]
\includegraphics[width=1.0\columnwidth]{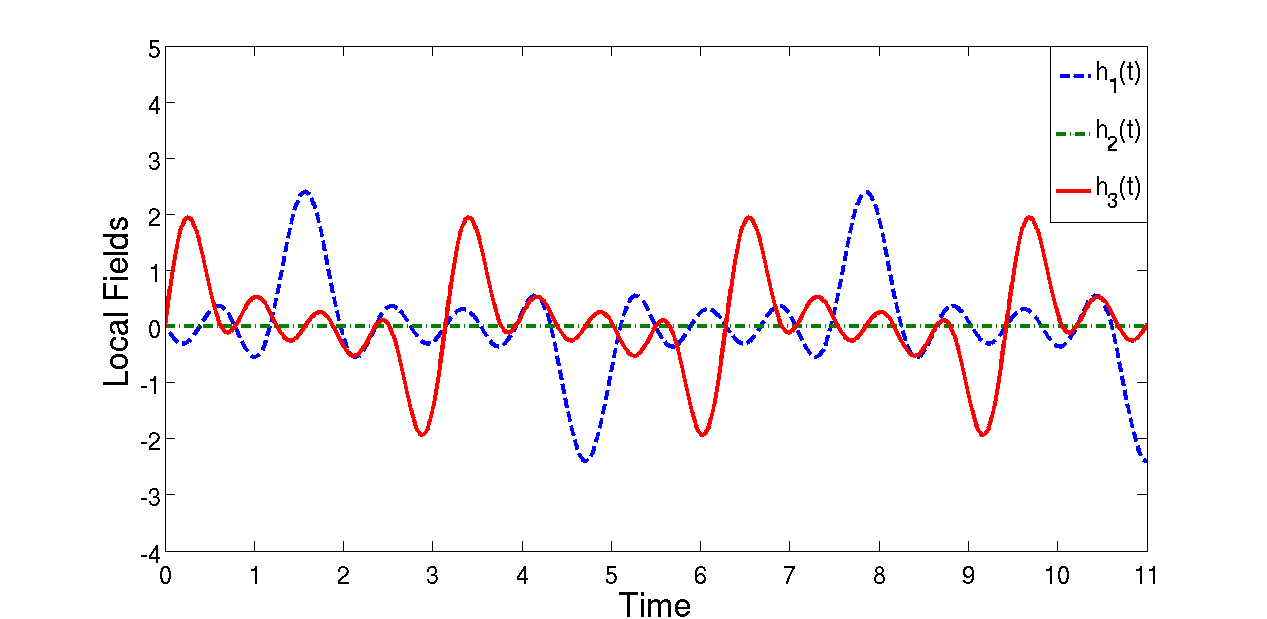} 
\includegraphics[width=1.0\columnwidth]{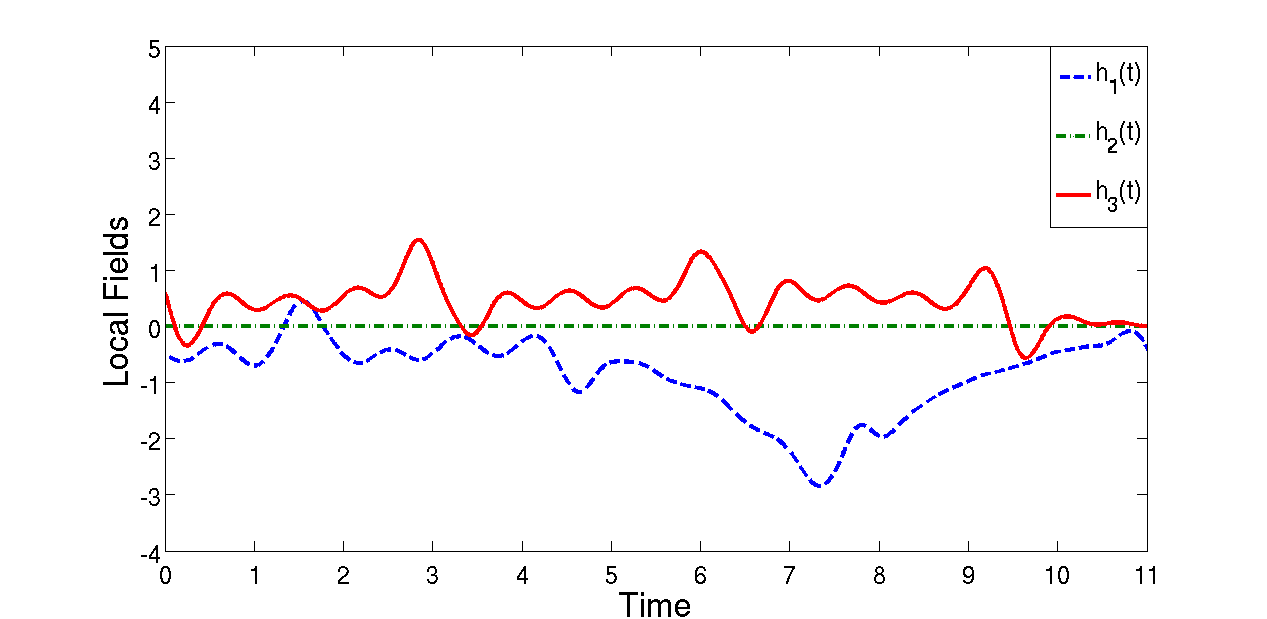}  \\ \hspace{-7.5cm} \text{b)} \hspace{8.2cm} \text{d)} \\ [-.05cm]
\includegraphics[width=1.0\columnwidth]{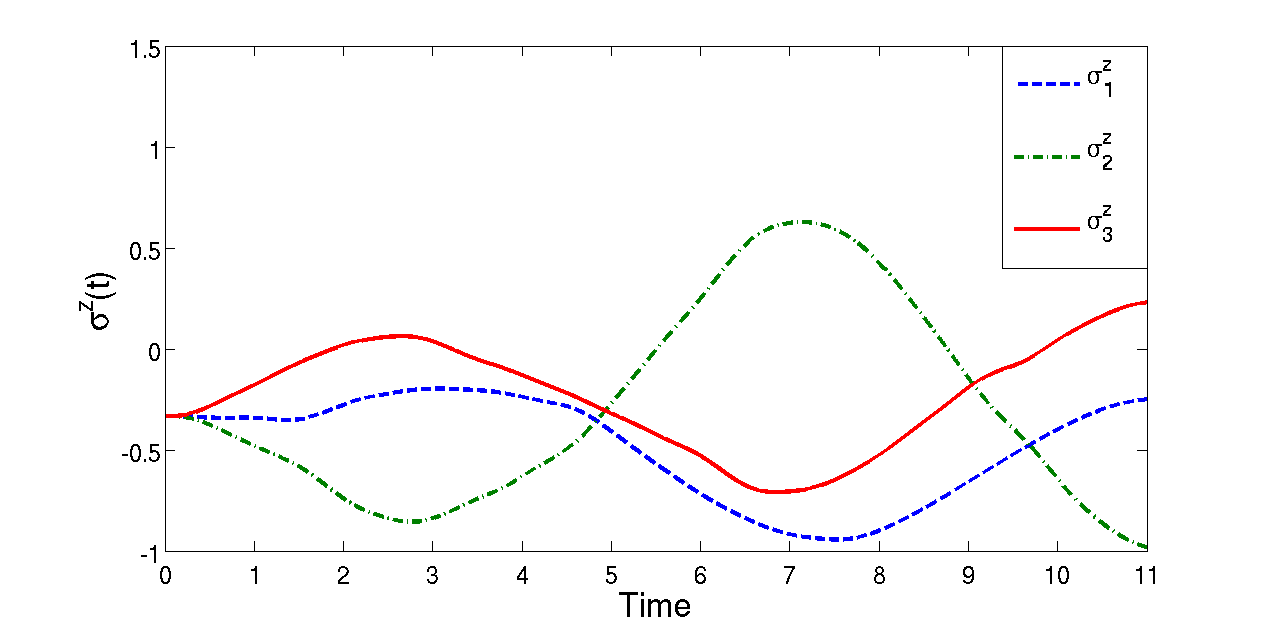} 
\includegraphics[width=1.0\columnwidth]{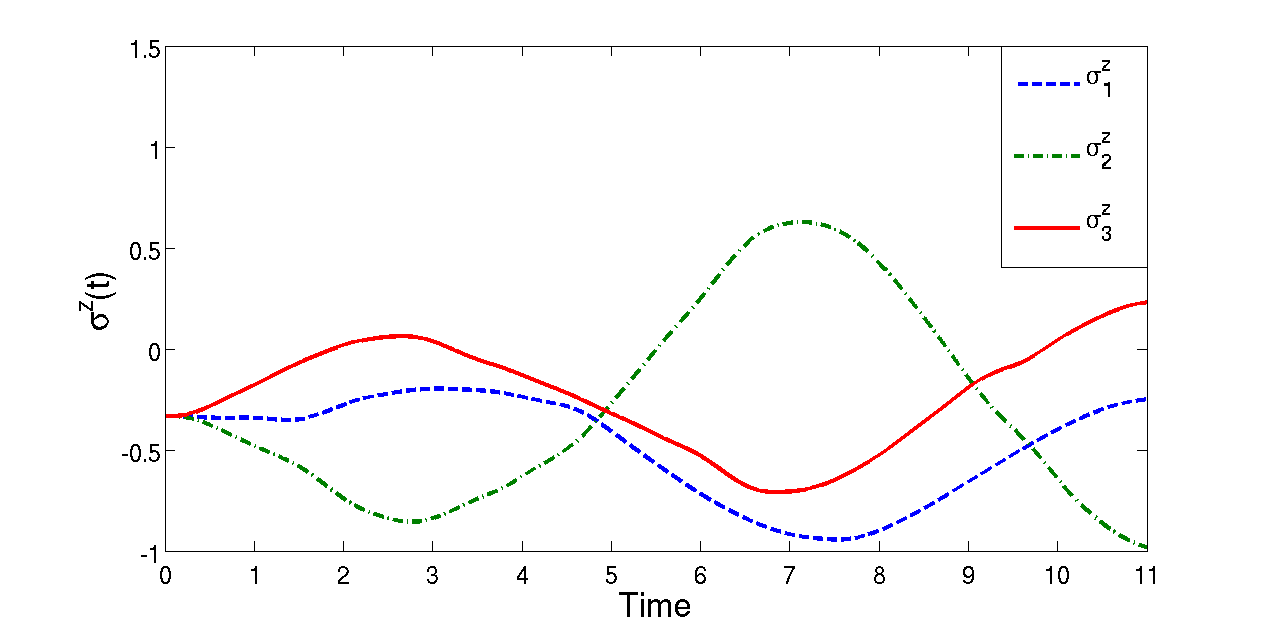} \\
\end{array}$
\end{centering}
\caption{\textbf{Simulating the Heisenberg Hamiltonain with the XY Hamiltonian} - Pulses of the form $h_1(t) = 0.6 \sum_{n=1}^{4}(-1)^{n+1} \rm{sin}\left[(2n-1)t\right]$ and $h_3(t) = 0.6 \sum_{n=1}^{4}(-1)^{2n} \rm{sin}\left[2nt\right]$ are respectively applied to the first and third qubits of a uniform Heisenberg Hamiltonian (a). The time-dependent Schr{\"o}dinger equation is then solved exactly numerically and the evolution of the set $\{ \sigma_1^z , \sigma_2^z ,\sigma_3^z  \}$ is read out in (b). The VL theorem gives us a prescription for constructing different auxiliary fields (c), which simulate the evolution of the set $\{ \sigma_1^z , \sigma_2^z ,\sigma_3^z  \}$ correctly as seen in (d), but using a non-uniform XY interaction instead. (Time is measured in units of $\frac{\hbar}{2J}$)}
\label{simulation_figure}
\end{figure*}

The proof of the VL theorem also gives a mathematical procedure (see supplementary material) for engineering the \textit{exact} auxiliary fields $\{ h'_1, h'_2, ... h'_N \}$ which reproduce a given set $\{ \sigma_1^z , \sigma_2^z , ... \sigma_N^z  \}$ under a different two-qubit interaction. As a simple demonstration, we use this procedure to numerically simulate a 3-qubit Heisenberg Hamiltonian using an XY Hamiltonian as the auxiliary system (Figure~\ref{simulation_figure}). For the simulation, the system is prepared in the initial state $|\psi(0)\rangle = \frac{1}{\sqrt{3}}(|011\rangle + |101\rangle + |110\rangle)$, where $|1 \rangle$ and $|0\rangle$ are eigenstates of $\hat{\sigma}^z$ with eigenvalues -1 and 1 respectively. In the Heisenberg Hamiltonian, $J^{\perp}_{i, i+1} = J^{\parallel}_{i, i+1} \equiv J_{i, i+1}$ and we choose $J_{12} = J_{23} = 0.5$. We apply a pulse of the form $h_1(t) = 0.6 \sum_{n=1}^{4}(-1)^{n+1} \rm{sin}\left[(2n-1)t\right]$ to the first qubit and $h_3(t) = 0.6 \sum_{n=1}^{4}(-1)^{2n} \rm{sin}\left[2nt\right]$ to the third qubit. The time-dependent Schr{\"o}dinger equation is solved numerically and the set $\{ \sigma_1^z , \sigma_2^z , \sigma_3^z  \}$ is read out during the evolution. Details of the simulation are provided in the supplementary material. For the auxiliary XY Hamiltonian, $J'^{\parallel}_{i, i+1} = 0$ and we choose different and non-uniform couplings in which $J'^{\perp}_{12} = 1.2$ and $J'^{\perp}_{23}= -1$. Using the VL theorem, we engineer the auxiliary local fields $\{ h'_1, h'_2, h'_3\}$ which using a non-uniform XY interaction, reproduce the set $\{ \sigma_1^z , \sigma_2^z , \sigma_3^z  \}$  obtained from the original evolution under the uniform Hesienberg Hamiltonian. As seen in Figure~\ref{simulation_figure}, the auxiliary local fields are quite different from the original local fields applied to the Heisenberg model, but simulate the set of components $\{ \sigma_1^z , \sigma_2^z , \sigma_3^z  \}$ correctly. i.e. $\{ \sigma_1'^z , \sigma_2'^z , \sigma_3'^z  \}$ = $\{ \sigma_1^z , \sigma_2^z , \sigma_3^z  \}$. In the language of electronic DFT, the XY model in our simulation is analogous to the "Kohn-Sham system" and the set $\{ h'_1, h'_2, h'_3\}$ play the role of the exact Kohn-Sham potential as a density functional. The VL theorem could also be useful from an experimental perspective, as it allows one to engineer different pulses which perform the same computations, but using different two-qubit couplings.

The RG and VL theorems place TDDFT for universal quantum computation on a firm theoretical footing and open several exciting research avenues. The development of approximate density functionals has been essential for the success of electronic DFT and TDDFT and will be in quantum computation and information theory as well. Functionals that perform actual computational tasks or subroutines such as the quantum Fourier transform are immediate research directions. The collapse of the computational complexity class hierarchy is of course not expected, and therefore finding functionals that carry out complex quantum computational tasks is extremely unlikely. Nevertheless, understanding how TDDFT functionals can {\sl approximate} quantum algorithms is an open direction. Density functionals for strongly correlated lattice and spin systems have been recently proposed~\cite{Verdozzi_Hubbard, Verdozzi_Hubbard_DMFT, Capelle_Europhys, Capelle_GS} and could be applied to several problems of relevance in quantum computing. In Refs.~\cite{Verdozzi_Hubbard, Verdozzi_Hubbard_DMFT, Capelle_Europhys, Capelle_GS} local density (LDA) and generalized gradient approximations (GGA) for one dimensional Hubbard chains and spin chains were derived from exact Bethe ansatz solutions and could readily be applied to solid-state quantum computing or perfect state transfer protocols in spin networks~\cite{state_transfer}. Functionals can also be parametrized from numerical simulations of one-dimensional qubit systems using time-dependent density matrix renormalization group methods (TDMRG)~\cite{TDMRG}, in an analogous fashion as quantum Monte Carlo simulations of the uniform electron gas have proven invaluable in electronic DFT~\cite{Monte_Carlo}. 

Another important research direction will be the generalization of DFT and TDDFT to other universal Hamiltonians and models of quantum computation. For instance, Ref.~\cite{DFT_AQC} discussed the use of TDDFT for obtaining gaps in adiabatic quantum computation. In Ref.~\cite{Lidar_Sham }, groundstate DFT was used to study relationships between entanglement and quantum phase transitions, while Ref.~\cite{Norbert_Frank} explored DFT from a complexity theory perspective. We are also exploring an extension of the TDDFT theorems to models with directly tunable two-qubit interactions, as opposed to the fixed interactions discussed in this letter.

Useful discussions with S. Mostame, J. D. Whitfield, S. Boxio, M. H. Yung and J. Parkhill are greatfully acknowledged. We thank NSF award PHY-0835713 for financial support.


\section*{Supplemental Material}

\subsection{Proof of the Runge-Gross theorem}

In this section we will first consider a proof of the RG theorem for Hamiltonians of the form,
\begin{equation}
\hat{H}(t) = \sum_{i < j}^N J^{\perp}_{ij} (\hat{\sigma}_i^{x} \hat{\sigma}_j^{x} + \hat{\sigma}_i^{y} \hat{\sigma}_j^{y}) + \sum_{i < j}^N J^{\parallel}_{ij} \hat{\sigma}_i^{z} \hat{\sigma}_j^{z} + \sum_{i=1}^N h_i(t) \hat{\sigma}_i^z,
\label{qubit Hamiltonian general}
\end{equation}
which reduces to Eq.~\ref{qubit Hamiltonian} of the main text in the limit of a one-dimensional array with nearest-neighbor couplings and open boundary conditions. We will see that it is possible to formulate the RG theorem of time-dependent current density functional theory (TDCDFT) for the more general class of Hamiltonians in Eq.~\ref{qubit Hamiltonian general}, but for TDDFT one must stay with the more restricted form in Eq.~\ref{qubit Hamiltonian}.

The proof begins with the equation of motion for the expectation value of the ith qubit along the field direction (z-axis),
\begin{equation}
\frac{\partial}{\partial t} \sigma_i^z = \imath \langle \left[\hat{H}(t), \hat{\sigma}_i^z \right] \rangle,
\label{commutator}
\end{equation}
where $\langle \hat{O} \rangle \equiv \langle \psi(t)| \hat{O} |\psi(t) \rangle$ denotes the expectation value of an arbitrary operator $\hat{O}$ at time t and $|\psi(t) \rangle$ is the wavefunction evolved on an interval [0,t] from a given initial state $|\psi(0) \rangle$, under the Hamiltonian in Eq.~\ref{qubit Hamiltonian general}. Development of the commutator in Eq.~\ref{commutator} yields,
\begin{eqnarray}
&& \frac{\partial}{\partial t} \sigma_i^z = 2 \sum_{k \neq i}^N J_{ki}^{\perp}( \langle \hat{\sigma}_k^x \hat{\sigma}_i^y \rangle - \langle \hat{\sigma}_k^y \hat{\sigma}_i^x \rangle ) \nonumber \\ &=& -\frac{1}{\imath} \sum_{k \neq i}^N J_{ki}^{\perp} ( \langle \hat{\sigma}_k^+ \hat{\sigma}_i^- \rangle - \langle \hat{\sigma}_k^- \hat{\sigma}_i^+ \rangle ),
\label{density_eom}
\end{eqnarray}
where we have introduced the Pauli raising and lowering operators $\hat{\sigma}^{\pm} = \hat{\sigma}^x \pm \imath \hat{\sigma}^y$ in the second equality. Defining
\begin{equation}
\hat{j}_{ki}  \equiv - 2  J_{ki}^{\perp}( \hat{\sigma}_k^x \hat{\sigma}_i^y - \hat{\sigma}_k^y \hat{\sigma}_i^x  ) = \frac{1}{\imath}J_{ki}^{\perp} ( \hat{\sigma}_k^+ \hat{\sigma}_i^-  -  \hat{\sigma}_k^- \hat{\sigma}_i^+  )
\label{current_operator}
\end{equation}
as the operator that generates the "current" of $\sigma^z$ flowing from the ith qubit to the kth qubit, Eq.~\ref{density_eom} takes the form of a local conservation law,
\begin{equation}
\frac{\partial}{\partial t} \sigma_i^z = - \sum_{k \neq i}^N \langle \Hat{j}_{ki} \rangle.
\label{conservation_law}
\end{equation}
This arises from the fact that the Hamiltonian in Eq.~\ref{qubit Hamiltonian general} conserves the total component of all N-qubits along the field direction. i.e. it is readily verified that,
\begin{equation}
\frac{\partial}{\partial t} \sum_i^N \sigma_i^z = \frac{\partial}{\partial t} \sigma_{total}^z = 0.
\end{equation}
This is analogous to the situation in electronic structure theory, where the local continuity equation
\begin{equation}
\frac{\partial}{\partial t} n(\mathbf{r}, t) = - \mathbf{\nabla} \cdot \mathbf{j}(\mathbf{r}, t)
\label{continuity}
\end{equation}
implies a global conservation of particle number
\begin{equation}
\frac{\partial}{\partial t} \int n(\mathbf{r}, t) d^3 \mathbf{r} = \frac{\partial}{\partial t} N = 0,
\end{equation}
where N is the number of electrons in the system.

We now consider a "primed" Hamiltonian
\begin{equation}
\hat{H}'(t) = \sum_{i < j}^N J^{\perp}_{ij} (\hat{\sigma}_i^{x} \hat{\sigma}_j^{x} + \hat{\sigma}_i^{y} \hat{\sigma}_j^{y}) + \sum_{i < j}^N J^{\parallel}_{ij} \hat{\sigma}_i^{z} \hat{\sigma}_j^{z} + \sum_{i=1}^N h'_i(t) \hat{\sigma}_i^z,
\label{primed qubit Hamiltonian RG}
\end{equation}
which has the same two-qubit interaction terms as the Hamiltonian in Eq.~\ref{qubit Hamiltonian general}, but a different set of local fields $\{h'_1, h'_2, ... h'_N \}$. Let $|\psi'(t) \rangle$ denote the wavefunction evolved from the \textit{same} initial state $|\psi (0) \rangle$, but under the Hamiltonian in Eq.~\ref{primed qubit Hamiltonian RG}. The equation of motion for the expectation value of the ith qubit along the z-axis under this primed Hamiltonian is
\begin{equation}
\frac{\partial}{\partial t} \sigma_i'^z = - \sum_{k \neq i}^N \langle \Hat{j}_{ki} \rangle',
\label{primed_conservation_law}
\end{equation}
where we define $\langle \hat{O} \rangle ' \equiv \langle \psi'(t)| \hat{O} |\psi'(t) \rangle$ as the expectation value of an arbitrary operator $\hat{O}$ with respect to the primed wavefunction. 



In what follows, we assume that the local fields in both the primed and unprimed systems are equal to their Taylor expansions within a finite radius of convergence around $t=0$. i.e. $h_i(t) = \sum_{j = 0}^{\infty} \left[ \frac{1}{j !} \frac{\partial^j}{\partial t^j} h_i(t) \right]_{t=0} t^j$ and similarly $h_i'(t) = \sum_{j = 0}^{\infty} \left[ \frac{1}{j !} \frac{\partial^j}{\partial t^j} h_i'(t) \right]_{t=0} t^j$ (the assumption of Taylor expandability is discussed below). We will now proceed to show that if the set of fields $\{ h_1, h_2, ... h_N \}$ differ from the set of fields $\{ h'_1, h'_2, ... h'_N \}$ by more than a global field which is the same for all N qubits, the set of currents $\{\langle \Hat{j}_{12} \rangle, \langle \Hat{j}_{13} \rangle, ... \langle \Hat{j}_{23} \rangle, ... \langle \Hat{j}_{N-1, N} \rangle \}$ and $\{\langle \Hat{j}_{12} \rangle', \langle \Hat{j}_{13} \rangle', ... \langle \Hat{j}_{23} \rangle', ... \langle \Hat{j}_{N-1, N} \rangle' \}$ will necessarily be different. The condition that the two sets of fields differ by more than a global field, is equivalent to the statement that there exists a smallest integer $m \geqslant 0$ such that the set
\begin{eqnarray}
&& \Big \{ \frac{\partial^m}{\partial t^m} (h_1(t) - h'_1(t))|_{t=0}, \frac{\partial^m}{\partial t^m} (h_2(t) - h'_2(t))|_{t=0}, \nonumber \\ [.3cm] &...& \frac{\partial^m}{\partial t^m} (h_N(t) - h'_N(t))|_{t=0} \Big \} \neq \{C\},
\label{condition}
\end{eqnarray}
where here $\{ C \}$ is a constant set of N elements that are all the same. i.e. the Taylor coefficients of the local fields in the primed and unprimed systems will differ at some order.

Next, we write down the equation of motion for the difference of the currents between the ith and kth qubits in the primed and unprimed systems:
\begin{equation}
\frac{\partial}{\partial t}(\langle \hat{j}_{ki} \rangle - \langle \hat{j}_{ki} \rangle') = \imath \langle \left[ \hat{H}(t), \hat{j}_{kl} ]\right \rangle - \imath \langle \left[ \hat{H}(t), \hat{j}_{kl} ]\right \rangle '.
\label{first_derivative_real-time}
\end{equation}
Since both systems evolve from a common initial state $|\psi(0) \rangle$, we have at $t=0$,
\begin{eqnarray}
&& \frac{\partial}{\partial t}(\langle \hat{j}_{ki} \rangle - \langle \hat{j}_{ki} \rangle')|_{t=0} = \imath \langle \psi(0) | \left[ (\hat{H}(0) - \hat{H}'(0)), \hat{j}_{ki} \right] |\psi(0) \rangle \nonumber \\  [.3cm] &=& 4 \langle \psi(0)| \hat{T}_{ki} | \psi(0) \rangle (\Delta h_{i}(0) - \Delta h_{k}(0) ).
\label{first_derivative}
\end{eqnarray} 
Here, we have defined $\Delta h_{i}(t) = h_i(t) - h'_i(t)$ as the difference between the unprimed and primed fields acting on the ith qubit and similarly, $\Delta h_{k}(t) = h_k(t) - h'_k(t)$. 
\begin{equation} 
\hat{T}_{ki} \equiv J^{\perp}_{k i} (\hat{\sigma}_k^{x} \hat{\sigma}_i^{x} + \hat{\sigma}_k^{y} \hat{\sigma}_i^{y}) = \frac{J^{\perp}_{k i}}{2}(\hat{\sigma}_k^{+} \hat{\sigma}_i^{-} + \hat{\sigma}_k^{-} \hat{\sigma}_i^{+})
\label{kinetic_operator}
\end{equation}
 is similar to a local kinetic energy operator, describing the total transfer of $\sigma^z$ between the ith and kth qubits. From Eq.~\ref{first_derivative}, we see that if the condition in Eq.~\ref{condition} is satisfied for $m=0$,  the sets $\{\langle \Hat{j}_{12} \rangle, \langle \Hat{j}_{13} \rangle, ... \langle \Hat{j}_{23} \rangle, ... \langle \Hat{j}_{N-1, N} \rangle \}$ and $\{\langle \Hat{j}_{12} \rangle', \langle \Hat{j}_{13} \rangle', ... \langle \Hat{j}_{23} \rangle', ... \langle \Hat{j}_{N-1, N} \rangle' \}$ will become different instantaneously later than $t=0$ (with a restriction on the vanishing of $\langle \psi(0)| \hat{T}_{ki} | \psi(0) \rangle$ discussed below). If the condition in Eq.~\ref{condition} instead holds for some $m > 0$, we differentiate Eq.~\ref{first_derivative_real-time} $m$ times to obtain, 
\begin{eqnarray}
&& \frac{\partial^{m+1}}{\partial t^{m+1}}(\langle \hat{j}_{ki} \rangle - \langle \hat{j}_{ki} \rangle')|_{t=0} \nonumber \\ &=&  4 \langle \psi(0)| \hat{T}_{ki} | \psi(0) \rangle \frac{\partial^{m}}{\partial t^{m}}( \Delta h_{i}(t) - \Delta h_{k}(t))|_{t=0}.
\label{kth_derivative}
\end{eqnarray}
From here we see that if the set of local fields eventually differ at any order, the set of currents must as well. This establishes the RG theorem of TDCDFT: For a fixed initial state $|\psi(0) \rangle$, there is a one to one mapping between the set of local fields and the set of currents, up to a globally constant field. 

We now discuss the three main conditions of the theorem: \\ [.3cm] 1) The expectation values $\langle \psi(0)| \hat{T}_{ki} | \psi(0) \rangle$ must be non-zero for at least one pair of qubits k and i, whose local field differences $\Delta h_{i}(t)$ and  $\Delta h_{k}(t)$ are different for at least one instant on the interval [0,t]. This is a fairly mild restriction on the set of admissible initial states, $|\psi(0) \rangle$. For instance, consider a worst case scenario, in which all the fields $\{ h_1, h_2, ... h_N \}$ and $\{ h'_1, h'_2, ... h_N \}$ differ by a constant field, except for $h_1$ and $h'_1$ which differ by a different amount from the others at only one instant in time on the interval [0,t]. In this worst case, the restriction means that $\langle \psi(0)| \hat{T}_{1i} | \psi(0) \rangle$ must be non-zero for at least one value of i, where $i = 1, 2, ..., N$. In the more general case, where the sets $\{ h_1, h_2, ... h_N \}$ and $\{ h'_1, h'_2, ... h_N \}$ differ for several qubits or on finite time intervals, this restriction is even less severe. \\ [.3cm] 2) The elements of the sets $\{ h_1, h_2, ... h_N \}$ and $\{ h'_1, h'_2, ... h_N \}$ must be analytic functions of time. i.e. equal to their Taylor expansions within a finite radius of convergence. In quantum computing, this is not a very severe restriction, as one typically constructs pulses which are well behaved functions. This restriction does not even exclude sudden switching, which is the case when applying idealized pulses to perform single-qubit rotations. \\[.3cm] 3) The theorem establishes a one to one mapping between  the set of currents and the set of local fields up to a globally constant field, we will denote $C(t)$, which is the same for all N qubits. If one applies periodic boundary conditions, the extra symmetry fixes the value of $C(t)$ and the mapping is one to one between the fields and currents with no ambiguity. For open boundary conditions, $C(t)$ remains arbitrary, which corresponds to an arbitrary term $C(t)\sum_i^N \hat{\sigma}_i^z \equiv C(t) \hat{\sigma}_{total}^z$ in the Hamiltonian. If one begins in an initial state which is an eigenstate of $\hat{\sigma}_{total}^z$, this term is simply a c-number and adds a trivial global phase to the wavefunction. This is typically the case when one begins in a computational basis state. However, if one starts in a superposition of states with different values of $\sigma_{total}^z$, the term C(t) $\hat{\sigma}_{total}^z$ yields a nontrivial coherence between these states. Such coherences would be measurable for observables with non-zero matrix elements between states of different $\sigma_{total}^z$. These observables would therefore not be uniquely determined by the current when considering open boundary conditions.

\begin{figure}
\begin{centering}
\includegraphics[width=1.0\columnwidth]{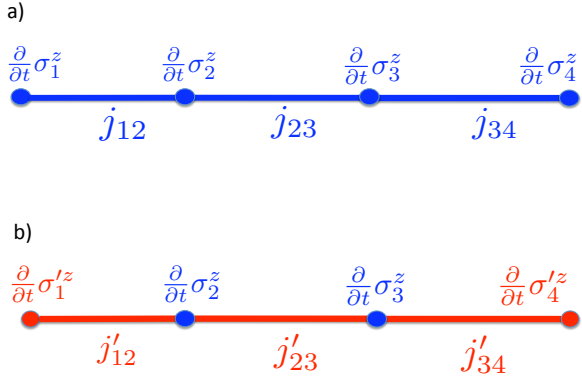} 
\end{centering}
\caption{\textbf{Open Chain of 4 qubits with nearest-neighbor couplings} - If we consider the set of currents shown in a), it will never be possible to find a new set of currents that will yield the same derivatives of $\sigma^z$. For instance, as shown in b), we can find a new set of currents that will yield the same derivatives of $\sigma^z$ on the middle two sites, but the derivatives at the ends of the chain will necessarily be different.}
\label{Open_Chain}
\end{figure}

\begin{figure*}[h]
\begin{centering}
$\begin{array}{c@{\hspace{1in}}c}
\includegraphics[width=1.0\columnwidth]{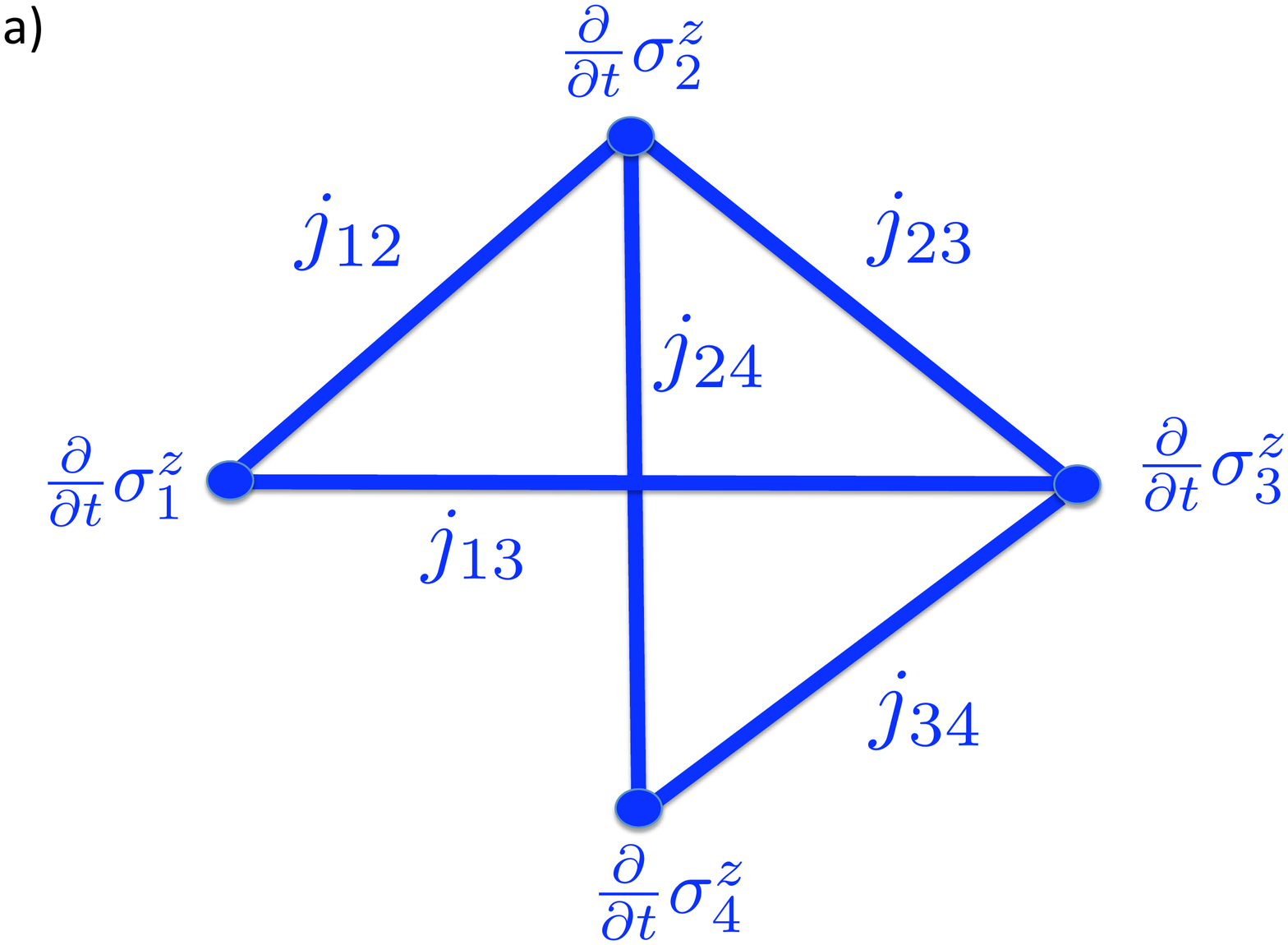} 
\includegraphics[width=1.0\columnwidth]{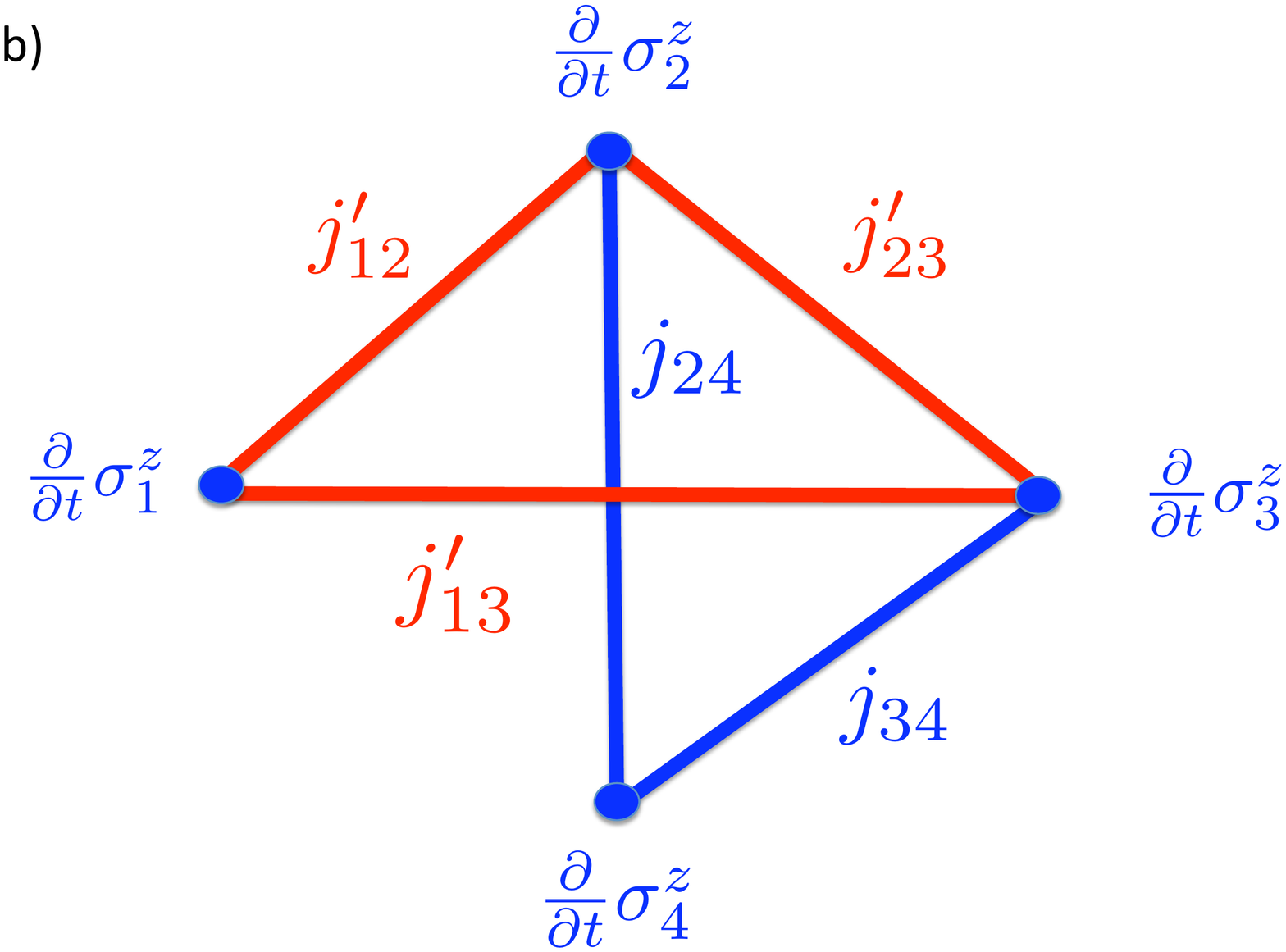}   \end{array}$
\end{centering}
\caption{\textbf{Open chain of 4 qubits with second nearest-neighbor couplings} - If we add second nearest neighbor couplings as well (shown in a)), it will be possible to find two sets of currents that yield the same derivatives of $\sigma^z$. For instance, if one considers a new set of currents around the closed loop shown in b), such that $\{ \delta j_{12}, \delta j_{13}, \delta j_{23}\}$ = constant, the derivatives of $\sigma^z$ will remain unchanged. This is equivalent to adding a purely transverse term to the current.}
\label{Second_Neighbor_Chain}
\end{figure*}

We now turn to the RG theorem of TDDFT, which is discussed in the main text. From Eq.~\ref{conservation_law}, we see that it is possible for two different sets of currents $\{\langle \Hat{j}_{12} \rangle, \langle \Hat{j}_{13} \rangle, ... \langle \Hat{j}_{23} \rangle, ... \langle \Hat{j}_{N-1, N} \rangle \}$ and $\{\langle \Hat{j}_{12} \rangle', \langle \Hat{j}_{13} \rangle', ... \langle \Hat{j}_{23} \rangle', ... \langle \Hat{j}_{N-1, N} \rangle' \}$ to correspond to the same set of spin components $\{ \sigma_1^z, \sigma_2^z,... \sigma_N^z\}$, if there exists a set of current differences
\begin{eqnarray}
&&  \{ \delta j_{12}, \delta j_{13}, ... \delta j_{23}, ... \delta j_{N-1, N} \} \nonumber \\ [.2cm] &\equiv& \{(\langle \Hat{j}_{12} \rangle - \langle \Hat{j}_{12} \rangle'), (\langle \Hat{j}_{13} \rangle -  \langle \Hat{j}_{13} \rangle'), \nonumber \\ [.15cm] &...& (\langle \Hat{j}_{23} \rangle - \langle \Hat{j}_{23} \rangle'), ... (\langle \Hat{j}_{N-1, N} \rangle - \langle \Hat{j}_{N-1, N} \rangle') \},
\end{eqnarray}
such that, 
\begin{equation}
\sum_{k \neq i}^N \delta j_{ik} = 0
\label{lattice_transverse}
\end{equation}
for some i. For a one-dimensional chain with open boundary conditions and nearest-neighbor couplings as in Eq.~\ref{qubit Hamiltonian}, such a set \textit{never} exists, as illustrated in Figure~\ref{Open_Chain}. Thus, for this case, no two sets of currents can yield the same set $\{ \sigma_1^z, \sigma_2^z,... \sigma_N^z\}$, and through the RG theorem of TDCDFT, no two sets of fields $\{ h_1, h_2, ... h_N \}$ and $\{ h'_1, h'_2, ... h'_N \}$ differing by more than a constant can yield the same set $\{ \sigma_1^z, \sigma_2^z,... \sigma_N^z\}$. This establishes the RG theorem of TDDFT for the Hamiltonian in Eq.~\ref{qubit Hamiltonian} of the text. For more general geometries, such as in Figure~\ref{Second_Neighbor_Chain}, it is possible to find two different sets of currents such that $\sum_{k \neq i}^N \delta j_{ik} = 0$. For these geometries, the RG theorem of TDCDFT holds, however that of TDDFT does not. From the continuity equation of electronic structure (Eq.~\ref{continuity}), we see that we can add an arbitrary transverse vector field $\vec{\delta j}$ (such that $\vec{\nabla} \cdot \vec{\delta j} = 0$) to the electronic current, without altering the value of $\frac{\partial}{\partial t} n(\mathbf{r}, t)$. A set $\{ \delta j_{12}, \delta j_{13}, ... \delta j_{23}, ... \delta j_{N-1, N} \}$ satisfying the condition in Eq.~\ref{lattice_transverse} is analogous to a purely transverse current.
\subsection{Proof of the van Leeuwen theorem}

The proof of the VL theorem begins with the equation of motion for the current under the evolution of the Hamiltonian in Eq.~\ref{qubit Hamiltonian general},
\begin{equation}
\frac{\partial}{\partial t} \langle \Hat{j}_{ki} \rangle = \imath \langle \left[\hat{H}(t), \Hat{j}_{ki} \right] \rangle.
\label{current_commutator}
\end{equation}
Developing the commutator in Eq.~\ref{current_commutator} yields,
\begin{equation}
\frac{\partial}{\partial t} \langle \Hat{j}_{ki} \rangle = \langle \hat{\mathscr T_{ki}} \rangle + \langle \hat{\mathscr F_{ki}} \rangle + 4 \langle \hat{T}_{ki} \rangle \delta h_{ik}(t),
\label{eom_current}
\end{equation}
where $\delta h_{ik}(t) = h_i(t) - h_k(t)$ is the difference between the local fields applied to the ith and kth qubits and we have defined the operators $\hat{\mathscr T_{ki}}$ and $\hat{\mathscr F_{ki}}$ as:
\begin{eqnarray}
&& \hat{\mathscr T_{ki}} \equiv 4  J_{ki}^{\perp} \Big\{ \hat{\sigma}_k^z  \left[\sum_{m \neq k} J_{mk}^{\perp} \hat{\sigma}_m^x \right] \hat{\sigma}_i^x + \hat{\sigma}_k^z  \left[\sum_{m \neq k} J_{mk}^{\perp} \hat{\sigma}_m^y \right] \hat{\sigma}_i^y \nonumber \\ &-& \hat{\sigma}_k^x  \left[\sum_{m \neq i} J_{mi}^{\perp} \hat{\sigma}_m^x \right] \hat{\sigma}_i^z - \hat{\sigma}_k^y  \left[\sum_{m \neq i} J_{mi}^{\perp} \hat{\sigma}_m^y \right] \hat{\sigma}_i^z \Big\},
\label{stress_tensor}
\end{eqnarray}
and
\begin{eqnarray}
&& \hat{\mathscr F_{ki}} \equiv 4  J_{ki}^{\perp} \Big\{ \hat{\sigma}_k^y  \left[\sum_{m \neq i} J_{mi}^{\parallel} \hat{\sigma}_m^z \right] \hat{\sigma}_i^y - \hat{\sigma}_k^x  \left[\sum_{m \neq i} J_{mi}^{\parallel} \hat{\sigma}_m^z \right] \hat{\sigma}_i^x \nonumber \\ &+&  \hat{\sigma}_k^y  \left[\sum_{m \neq k} J_{mk}^{\parallel} \hat{\sigma}_m^z \right] \hat{\sigma}_i^y - \hat{\sigma}_k^x  \left[\sum_{m \neq k} J_{mk}^{\parallel} \hat{\sigma}_m^z \right] \hat{\sigma}_i^x \Big\}.
\label{force_density}
\end{eqnarray}
$\hat{\mathscr T_{ki}}$ arises from the commutator of the current operator with the kinetic energy operator $\sum_{i<j}^N \hat{T}_{ij}$ and is similar to the "stress tensor" operator of electronic TDDFT~\cite{van leeuwen}. $\hat{\mathscr F_{ki}}$ arises from the commutator of the current operator with the term $\sum_{i < j}^N J^{\parallel}_{ij} \hat{\sigma}_i^{z} \hat{\sigma}_j^{z}$ in the Hamiltonian and is analogous to the "internal force density" operator due to the electron-electron repulsion in electronic TDDFT. Both the terms $\langle \hat{\mathscr T_{ki}} \rangle$ and $\langle \hat{\mathscr F_{ki}} \rangle$ in Eq.~\ref{eom_current} represent "internal forces" due to the two-qubit terms in the Hamiltonian, while the term $4 \langle \hat{T}_{ki} \rangle \delta h_{ik}(t)$ represents an "external" driving force due to the applied local fields which couple to the one-qubit operators $\hat{\sigma}_i^z$.

We now consider an auxiliary "primed" system, with the Hamiltonian,
\begin{equation}
\hat{H}'(t) = \sum_{i < j}^N J'^{\perp}_{ij} (\hat{\sigma}_i^{x} \hat{\sigma}_j^{x} + \hat{\sigma}_i^{y} \hat{\sigma}_j^{y}) + \sum_{i < j}^N J'^{\parallel}_{ij} \hat{\sigma}_i^{z} \hat{\sigma}_j^{z} + \sum_{i=1}^N h'_i(t) \hat{\sigma}_i^z,
\label{primed qubit Hamiltonian}
\end{equation}
having different two-qubit interactions $J'^{\perp}_{ij}$ and $J'^{\parallel}_{ij}$ and a different set of local fields $\{ h'_1, h'_2, ... h'_N \}$. We allow the initial state $|\psi'(0) \rangle$ to be different from the initial state $| \psi(0) \rangle$, with the only constraint that the initial currents and the initial expectation values of $\hat{\sigma}_i^z$ must be the same in the primed and unprimed systems. i.e. we only require that, 
\begin{equation}
\langle \psi'(0)| \Hat{j}_{ki}' |\psi'(0) \rangle = \langle \psi(0)| \Hat{j}_{ki} |\psi(0) \rangle
\label{current_condition}
\end{equation}
and 
\begin{equation}
\langle \psi'(0)| \hat{\sigma}_i^z |\psi'(0) \rangle = \langle \psi(0)| \hat{\sigma}_i^z |\psi(0) \rangle
\label{density_condition}
\end{equation}
 for all k and i. The equation of motion for the current under evolution of this Hamiltonian is,
\begin{equation}
\frac{\partial}{\partial t} \langle \Hat{j}_{ki}' \rangle' = \langle \hat{\mathscr T'_{ki}} \rangle' + \langle \hat{\mathscr F'_{ki}} \rangle' + 4 \langle \hat{T}'_{ki} \rangle' \delta h'_{ik}.
\label{eom_current_primed}
\end{equation}
Here, the operators $\Hat{j}_{ki}'$, $\hat{T}'_{ki}$,  $\hat{\mathscr T'_{ki}}$ and $\hat{\mathscr F'_{ki}}$ are defined exactly as for the unprimed system in Eq's~\ref{current_operator}, \ref{kinetic_operator}, \ref{stress_tensor} and \ref{force_density} respectively, but with $J^{\perp}_{ij}$ and $J^{\parallel}_{ij}$ replaced by $J'^{\perp}_{ij}$ and $J'^{\parallel}_{ij}$. 

We now assume that \textit{all} quantities in Eq.'s~\ref{eom_current} and~\ref{eom_current_primed} are equal to their Taylor series expansions with a finite radius of convergence around $t=0$. Denoting the Taylor expansion of an arbitrary function $f(t)$ as $f(t) = \sum_{i=0}^{\infty} f^m t^m$ and Taylor expanding both sides of Eq.~\ref{eom_current}, we find after equating coefficients of equal powers of t,
\begin{equation}
(m+1) \langle \Hat{j}_{ki} \rangle^{m+1} = \langle \hat{\mathscr T_{ki}} \rangle^{m} + \langle \hat{\mathscr F_{ki}} \rangle^{m} + 4 \sum_{s=0}^{m} \langle \hat{T}_{ki} \rangle^{ m-s} \delta h_{ik} ^{ s}. 
\label{eom_current_expanded}
\end{equation}
Similarly, for Eq.~\ref{eom_current_primed} we have,
\begin{equation}
(m+1) \langle \Hat{j}_{ki}' \rangle^{'m+1} = \langle \hat{\mathscr T'_{ki}} \rangle^{'m} + \langle \hat{\mathscr F'_{ki}} \rangle^{'m} + 4 \sum_{s=0}^{m} \langle \hat{T}'_{ki} \rangle^{' m-s} \delta h_{ik} ^{' s}. 
\label{eom_current_primed_expanded}
\end{equation}
We now subtract Eq.~\ref{eom_current_primed_expanded} from Eq.~\ref{eom_current_expanded} and \textit{demand} that the set of currents $\{\langle \Hat{j}_{12} \rangle, \langle \Hat{j}_{13} \rangle, ... \langle \Hat{j}_{23} \rangle, ... \langle \Hat{j}_{N-1, N} \rangle \}$ and $\{\langle \Hat{j}'_{12} \rangle', \langle \Hat{j}'_{13} \rangle', ... \langle \Hat{j}'_{23} \rangle', ... \langle \Hat{j}'_{N-1, N} \rangle' \}$ be the same in the primed and unprimed systems. i.e. we demand that the Taylor coefficients $\langle \Hat{j}_{ki} \rangle^{m}$ and $\langle \Hat{j}_{ki}' \rangle^{'m}$ be the same for all m and for all qubit pairs k and i. This yields
\begin{eqnarray}
&&  4 \langle \hat{T}'_{ki} \rangle^{' 0} \delta h_{ik} ^{' m} = -  4 \sum_{s=0}^{m-1} \langle \hat{T}'_{ki} \rangle^{' m-s} \delta h_{ik} ^{' s} +   4 \sum_{s=0}^{m} \langle \hat{T}_{ki} \rangle^{ m-s} \delta h_{ik} ^{ s} \nonumber \\ &+& \langle \hat{\mathscr T_{ki}} \rangle^{m} - \langle \hat{\mathscr T'_{ki}} \rangle^{' m}  +\langle \hat{\mathscr F_{ki}} \rangle^{m} - \langle \hat{\mathscr F'_{ki}} \rangle^{' m},
\label{recursion}
\end{eqnarray}
for all k and i. We see that the left hand side of Eq.~\ref{recursion} contains Taylor coefficients of $\delta h'_{ik}(t)$ of order m, while the right hand side has only Taylor coefficients of $\delta h'_{ik}(t)$ of order less than m and known quantities. Thus, when supplemented with the condition in Eq.~\ref{current_condition}, Eq.~\ref{recursion} is a unique recursion relation for the Taylor coefficients of the local field differences $\delta h'_{ik}(t)$, which reproduce the given set of currents $\{\langle \Hat{j}_{12} \rangle, \langle \Hat{j}_{13} \rangle, ... \langle \Hat{j}_{23} \rangle, ... \langle \Hat{j}_{N-1, N} \rangle \}$ using different two-qubit interactions $J'^{\perp}_{ij}$ and $J'^{\parallel}_{ij}$.  Eq.~\ref{recursion} can be used to construct the set local fields, $\{ h'_1, h'_2, ... h'_N \}$ up to an arbitrary constant field. For periodic boundary conditions, the arbitrary constant is again fixed by the extra symmetry and the fields themselves are uniquely determined.

So far we have established a VL theorem for TDCDFT. In order to establish the VL theorem of TDDFT discussed in the text, we simply need to add the condition in Eq.~\ref{density_condition}. With this additional constraint, it is clear from Eq's.~\ref{conservation_law} and~\ref{primed_conservation_law} that if the constructed fields $\{ h'_1, h'_2, ... h'_N \}$ force the set $\{\langle \Hat{j}'_{12} \rangle', \langle \Hat{j}'_{13} \rangle', ... \langle \Hat{j}'_{23} \rangle', ... \langle \Hat{j}'_{N-1, N} \rangle' \}$ to be the same as $\{\langle \Hat{j}_{12} \rangle, \langle \Hat{j}_{13} \rangle, ... \langle \Hat{j}_{23} \rangle, ... \langle \Hat{j}_{N-1, N} \rangle \}$, the sets $\{ \sigma_1'^z, \sigma_2'^z, ... \sigma_N'^z\}$ and $\{ \sigma_1^z , \sigma_2^z , ... \sigma_N^z  \}$ must be the same as well.

We now discuss two main conditions of the theorem: \\ [.3cm] 1) \textit{All} of the quantities appearing in Eq.'s~\ref{eom_current} and~\ref{eom_current_primed} as well as the sets $\{ \sigma_1^z, \sigma_2^z, ... \sigma_N^z\}$ and $\{ \sigma_1'^z , \sigma_2'^z , ... \sigma_N'^z  \}$ must be equal to their Taylor expansions within a finite radius of convergence for the theorem to hold. This is a much more restrictive condition than for the RG theorem, which only requires that the sets of fields $\{ h_1, h_2, ... h_N \}$ and $\{ h'_1, h'_2, ... h'_N \}$  be equal to their Taylor series expansions. This restriction arises in the VL theorem of electronic TDDFT as well and approaches to circumvent this condition have begun to be researched~\cite{VL_extension}. \\ [.3cm] 2) For the entire set of fields  $\{ h'_1, h'_2, ... h'_N \}$ to exist, $\langle \hat{T}'_{ki} \rangle^{' 0}$ must be non-vanishing for \textit{all} pairs of qubits k and l. This too is a more severe restriction on the class of admissible initial states than in the R.G. theorem, which only required that $\langle \hat{T}_{ki} \rangle^{0}$ be non-vanishing for certain values of k and l. However, since we are free to choose $|\psi'(0) \rangle$ so long as it satisfies the conditions in Eq.'s~\ref{current_condition} and~\ref{density_condition}, we will often be able to choose an initial state such that $\langle \hat{T}'_{ki} \rangle^{' 0} \neq 0$. From a practical standpoint, we have also found in our numerical simulations that for vanishing $\langle \hat{T}'_{ki} \rangle^{' 0}$, we can add a small convergence factor to make the fields well behaved at the initial time with little error in the overall propagation. This situation does not arise in electronic TDDFT for continuous systems, but similar problems have been noticed when one defines electronic TDDFT for lattice systems~\cite{Baer_Lattice, Ullrich_Lattice, pearls_phases, lattice_TDCDFT}. Since the qubit Hamiltonians we consider in this letter are also discrete, it is not surprising that a similar situation arises.

\subsection{Entanglement as a functional of the set $\{ \sigma_1^z, \sigma_2^z, ... \sigma_N^z \}$.}

In this section, we will discuss the construction of two-qubit entanglement as a functional of the single-qubit expectation values, $\{ \sigma_1^z, \sigma_2^z, ... \sigma_N^z \}$. We use the concurrence as a measure of the entanglement between any two qubits in an N-qubit system~\cite{Wooters}. Since the concurrence depends on non-commuting two-qubit observables, we expect that it is in general very hard to construct exactly as a functional of the set $\{ \sigma_1^z, \sigma_2^z, ... \sigma_N^z \}$, which are simple expectation values of commuting single-qubit observables. We will see that this is indeed the case, but in the spirit of electronic TDDFT, one can hope to develop simple approximations.

The two-qubit reduced density matrix (2RDM) for the kth and lth qubits is obtained by tracing the full N-qubit density matrix over all other $N-2$ qubits in the system. In this letter we consider only pure states, so the 2RDM is simply given by
\begin{equation}
\rho_{kl} =  \rm{Tr}_{1, ...k-1, k+1, ...l-1, l+1, ... N}\left[|\psi(t) \rangle \langle \psi(t)| \right],
\label{2RDM}
\end{equation}
where Tr denotes a partial trace. Defining the "time-reversed" 2RDM as
\begin{equation}
\tilde{\rho}_{kl} = \hat{\sigma}_k^y \otimes \hat{\sigma}_l^y \rho_{kl}^* \hat{\sigma}_k^y \otimes \hat{\sigma}_l^y,
\label{2RDM_time-reversed}
\end{equation}
the concurrence $E_{kl}$ is defined in terms of the eigenvalues $\lambda_i$ of the matrix $\rho_{kl} \tilde{\rho}_{kl}$ as,
\begin{equation}
E_{kl} = \rm{max} (0, \sqrt{\lambda_1} - \sqrt{\lambda_2} - \sqrt{\lambda_3} - \sqrt{\lambda_4}).
\label{concurrence}
\end{equation}
In Eq.~\ref{concurrence}, the eigenvalues $\lambda_i$ are arranged in decreasing order. i.e. $\lambda_1 > \lambda_2 > \lambda_3 > \lambda_4$.


We first investigate the concurrence for a system which is restricted to the $\sigma_{total}^z = \pm (N-2)$ subspace. There is only one flipped qubit relative to the other $N-1$ qubits. This is also known as the single-excitation manifold. We denote $|i\rangle = |00...010...00\rangle$ as the computational basis state with the ith qubit in the state $|1\rangle$ and all other qubits in the state $|0\rangle$ (the 0's and 1's can be interchanged without changing any results). The N-qubit density matrix can be expanded in terms of the N computational basis functions as,
\begin{equation}
|\psi(t) \rangle \langle \psi(t) | = \sum_{i,j =1}^{N} a_i^*(t) a_j(t) |j \rangle \langle i|.
\end{equation}
From the above expression, we find the 2RDM for the kth and lth qubits to be
\begin{equation}
\rho_{kl} = \left(\begin{array}{cccc}\sum_{i \neq k, l} |a_i(t)|^2 & 0 & 0 & 0 \\0 & |a_l(t)|^2 & a_l(t) a_k^* & 0 \\0 & a_l^*(t) a_k & |a_k(t)|^2 & 0 \\0 & 0 & 0 & 0\end{array}\right).
\label{2RDM_matrix}
\end{equation}
In Eq.~\ref{2RDM_matrix},  $\rho_{kl}$ is expressed in the 2-qubit computational basis states, $\{ |00 \rangle, |01 \rangle, |10 \rangle, |11 \rangle \}$. From Eq.'s~\ref{2RDM_time-reversed} and~\ref{concurrence}, we find the concurrence to be,
\begin{equation}
E_{kl} =2 |a_l(t)| |a_k(t)|.
\label{one_magnon_concurrence}
\end{equation}
In order to re-express $E_{kl}$ in terms of $\{ \sigma_1^z, \sigma_2^z, ... \sigma_N^z \}$, we need to invert the matrix equation
\begin{equation}
\vec{\sigma^z} = \overleftrightarrow{M} \vec{a},
\end{equation}
where $\vec{\sigma^z}$ and $\vec{a}$ are column vectors formed from the sets $\{ \sigma_1^z, \sigma_2^z, ... \sigma_N^z \}$ and $\{ |a_1|^2, |a_2|^2, ...|a_N|^2 \}$ respectively and $\overleftrightarrow{M}$ is a square matrix with the diagonal elements equal to -1 and all other elements equal to 1. Carrying out the inversion and substituting the result into Eq.~\ref{one_magnon_concurrence} yields Eq.~\ref{entanglement} of the main text. Thus, we see that in the case of a single flipped qubit, it is very simple to obtain an exact entanglement functional.

As a more complicated example, we consider entanglement in the $\sigma_{total}^z = \pm (N-4)$ subspace, which contains states with two flipped qubits. We denote $|ij\rangle = |00...010...010...00\rangle$ as the computational basis state with the ith and jth qubits in the state $|1\rangle$ and all other qubits in the state $|0\rangle$. Here, we can expand the N-qubit density matrix in terms of these $\frac{N(N-1)}{2}$ computational basis functions as,
\begin{equation}
|\psi(t) \rangle \langle \psi(t) | = \sum_{i < j, k <l } a_{ij}^*(t) a_{kl}(t) |kl \rangle \langle ij|.
\
\end{equation}
Obtaining the 2RDM as before, we find the eigenvalues of $\rho_{kl} \tilde{\rho}_{kl}$ to be
\begin{equation}
\lambda_1 = \Big\{ \sqrt{(\sum_{i \neq k, l} |a_{il}(t)|^2)(\sum_{j \neq k, l} |a_{jk}(t)|^2)} + |\sum_{i \neq k,l} a_{ki}(t) a^*_{li}(t)| \Big\}^2,
\end{equation}
\begin{equation}
\lambda_2 = \lambda_3 = |a_{kl}(t)|^2 \sum_{i < j \neq k,l} |a_{ij}(t)|^2,
\end{equation}
and
\begin{equation}
\lambda_4 = \Big\{ \sqrt{(\sum_{i \neq k, l} |a_{il}(t)|^2)(\sum_{j \neq k, l} |a_{jk}(t)|^2)} - |\sum_{i \neq k,l} a_{ki}(t) a^*_{li}(t)| \Big\}^2.
\end{equation}
The terms depending on sums over the coefficients' moduli squared, $|a_{ij}(t)|^2$, can be obtained fairly easily in terms of $\{ \sigma_1^z, \sigma_2^z, ... \sigma_N^z \}$ using the expression
\begin{equation}
\sigma_i^z = 1 -  \sum_{l \neq i}  |a_{il}(t)|^2.
\end{equation}
However, we see that $\lambda_1$ and $\lambda_4$ also contain the term $|\sum_{i \neq k,l} a_{ki}(t) a^*_{li}(t)|$, which depends explicitly on phases in the wavefunction. The imaginary parts of the coherences can be obtained from the currents, which in turn can be obtained from time derivatives of $\{ \sigma_1^z, \sigma_2^z, ... \sigma_N^z \}$. However, the real parts of the coherences depend on expectation values of the kinetic energy operators, $\hat{T}_{kl}$, and in general will need to be approximated. Since the number of computational basis states increases with the number of flipped qubits, we expect the exact entanglement functional to become progressively more complicated as more qubits are flipped. This highlights the need for constructing simple approximate entanglement functionals and will be explored in future work.

\subsection{Numerical propagation of the VL construction.}

\begin{figure*}[h]
\begin{centering}
$\begin{array}{c@{\hspace{1in}}c}
 \\[-0.53cm]  \text{\underline{\Large{Heisenberg interaction}} \hspace{5.0cm} \underline{\Large{XY interaction}}} \\ [0.3cm]\\ \hspace{-15cm} \text{a)} \\ [-.05cm]
\includegraphics[width=1.0\columnwidth]{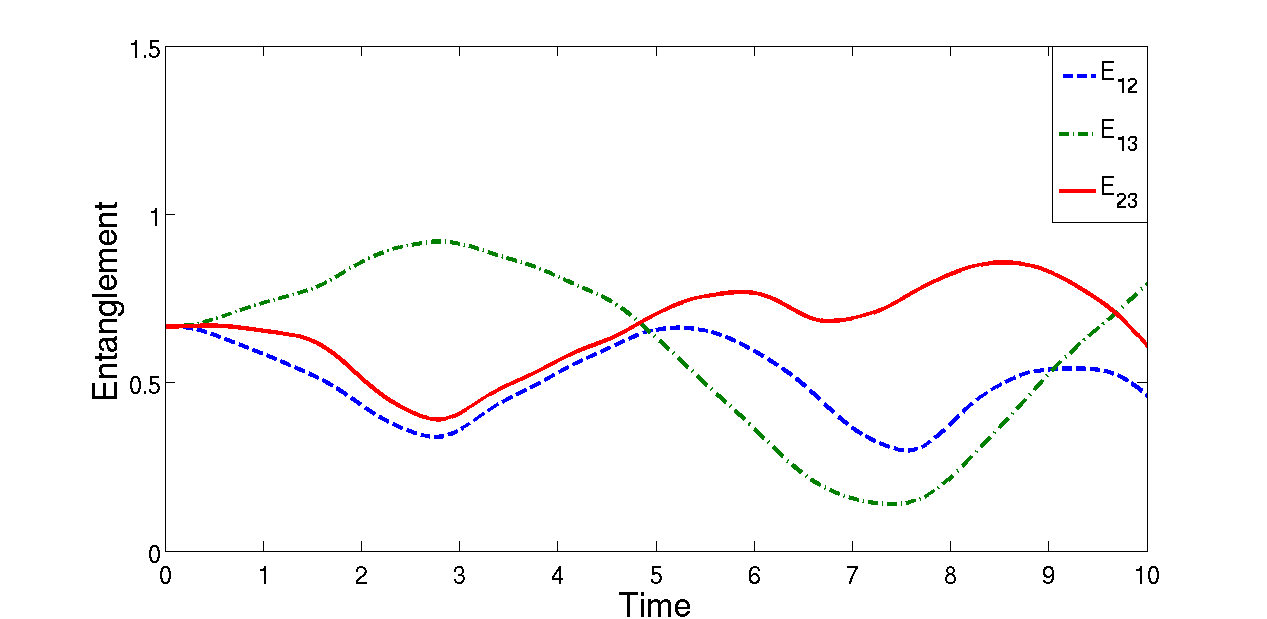} 
\includegraphics[width=1.0\columnwidth]{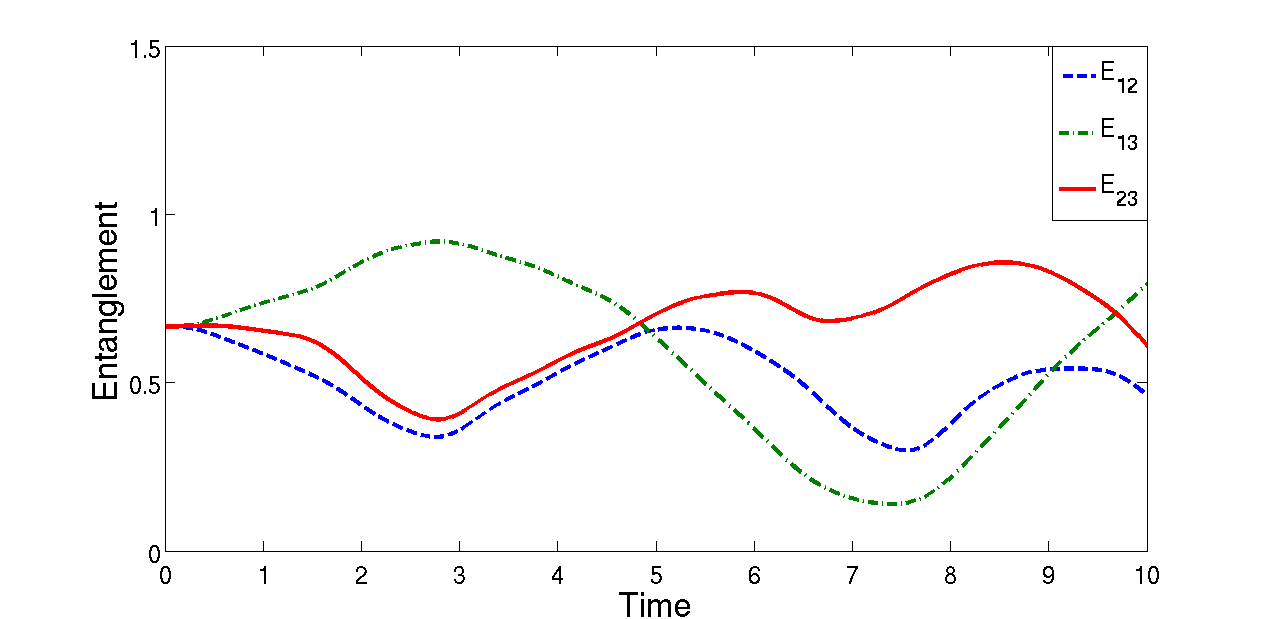}  \\ \hspace{-15cm} \text{b)}  \\ [-.05cm]
\includegraphics[width=1.0\columnwidth]{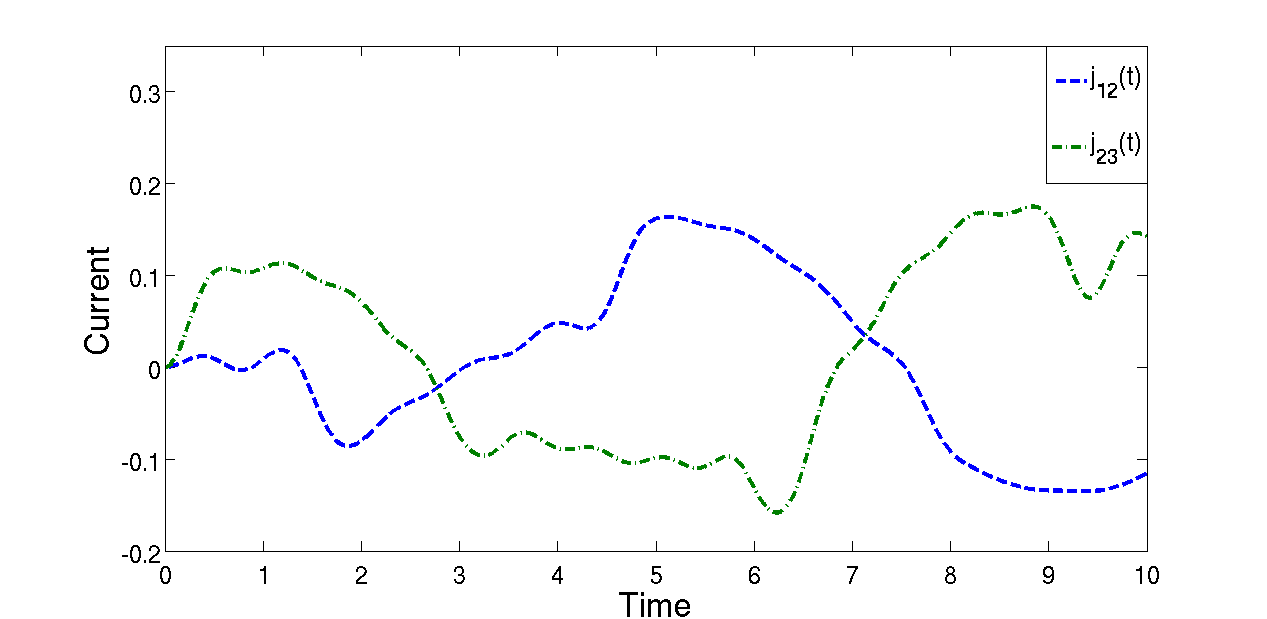} 
\includegraphics[width=1.0\columnwidth]{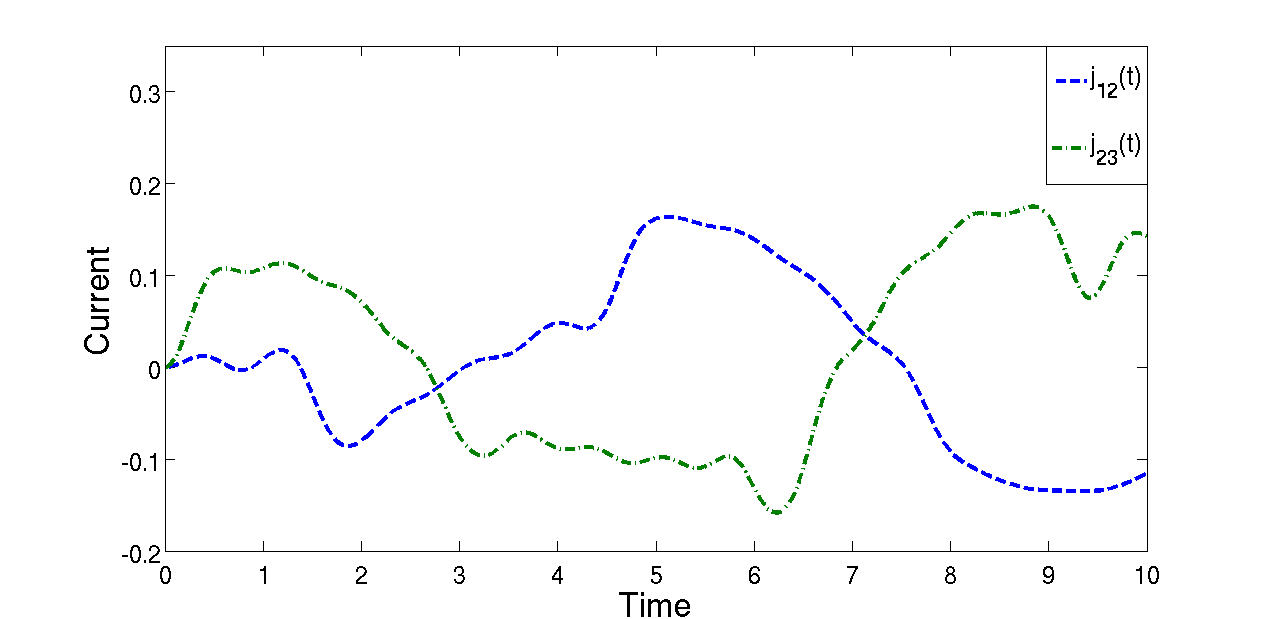} \\
\\ \hspace{-15cm} \text{c)}  \\ [-.05cm]
\includegraphics[width=1.0\columnwidth]{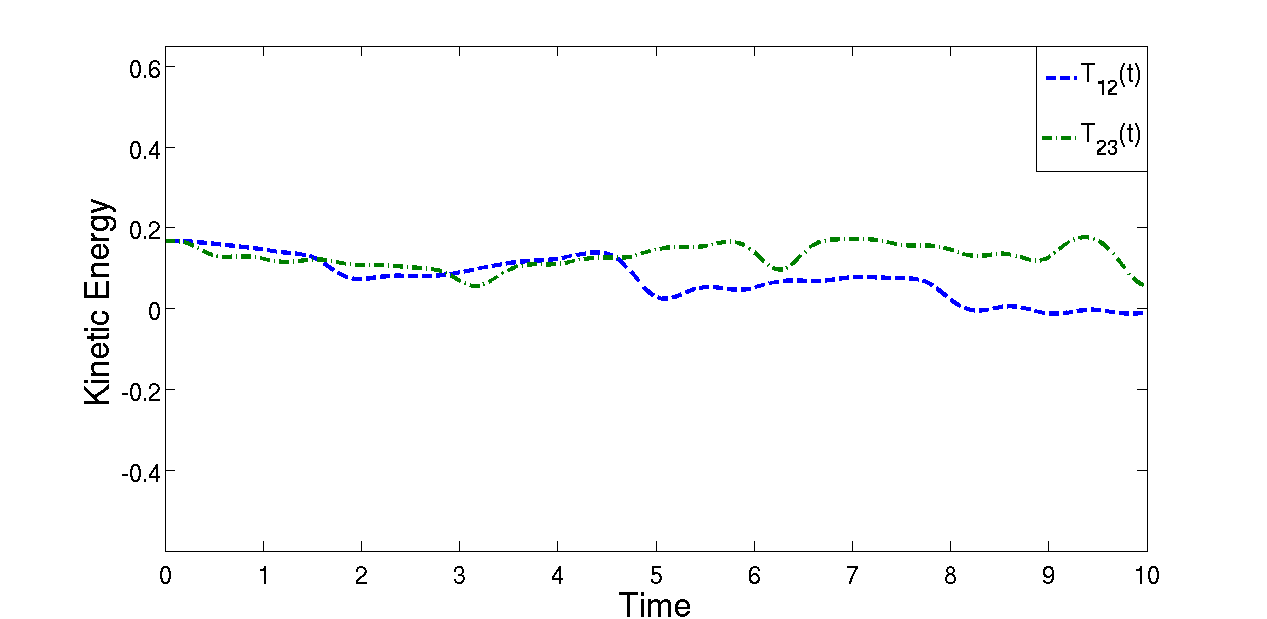} 
\includegraphics[width=1.0\columnwidth]{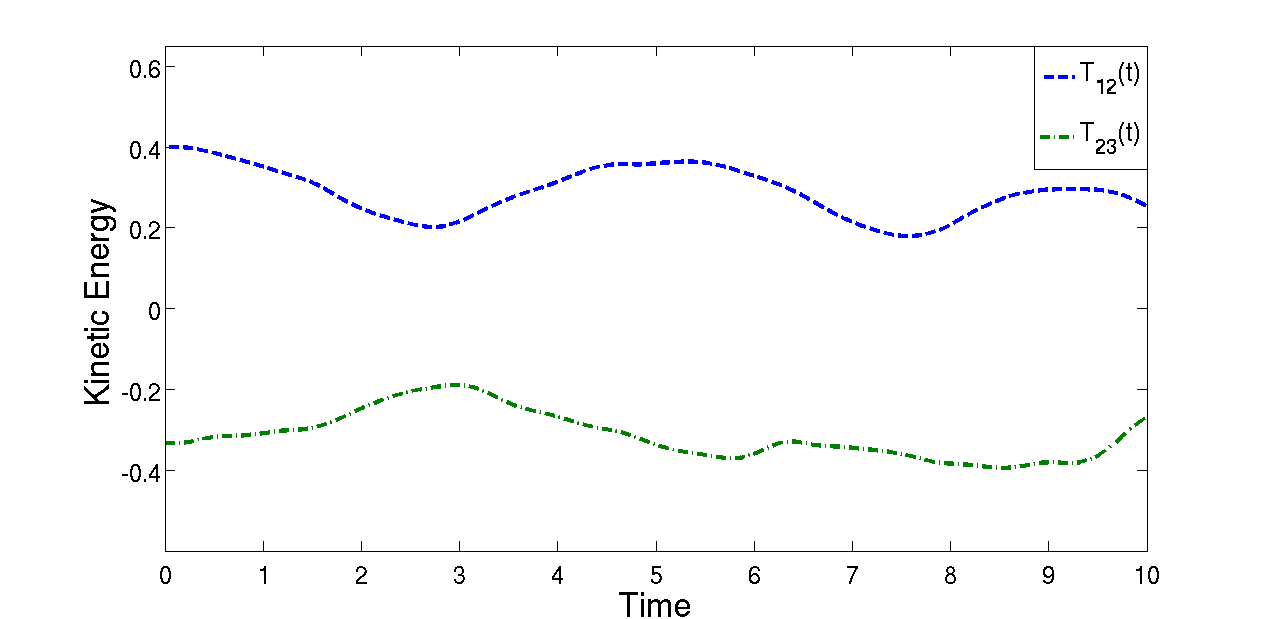}
\end{array}$
\end{centering}
\caption{\textbf{Observables in the Heisenberg Hamiltonian verses the XY Hamiltonian} - a) The entanglement calculated directly with the wavefunctions $|\psi(t) \rangle$ (left column) and $|\psi'(t) \rangle$ (right column) is seen to be the same, as expected since the propagation is restricted to the subspace with one flipped qubit and Eq.~\ref{entanglement} holds. b) The currents $\{\langle \Hat{j}_{12} \rangle, \langle \Hat{j}_{23} \rangle \}$ (left column) and $\{\langle \Hat{j}'_{12} \rangle', \langle \Hat{j}'_{23} \rangle' \}$ (right column) are also the same, as a consequence of the VL construction. c) The kinetic energies $\{\langle \hat{T}_{12} \rangle, \langle \hat{T}_{23} \rangle \}$ (left column) and $\{\langle \hat{T}'_{12} \rangle', \langle \Hat{T}'_{23} \rangle' \}$ (right column) are very different, as is often the case in electronic TDDFT as well.}
\label{simulation_figure_2}
\end{figure*}

\begin{figure*}[h]
\begin{centering}
$\begin{array}{c@{\hspace{1in}}c}
 \\[-0.53cm]  \text{\underline{\Large{Heisenberg interaction}} \hspace{5.0cm} \underline{\Large{XY interaction}}} \\ [0.3cm]\\ \hspace{-15cm} \text{a)} \\ [-.05cm]
\includegraphics[width=1.0\columnwidth]{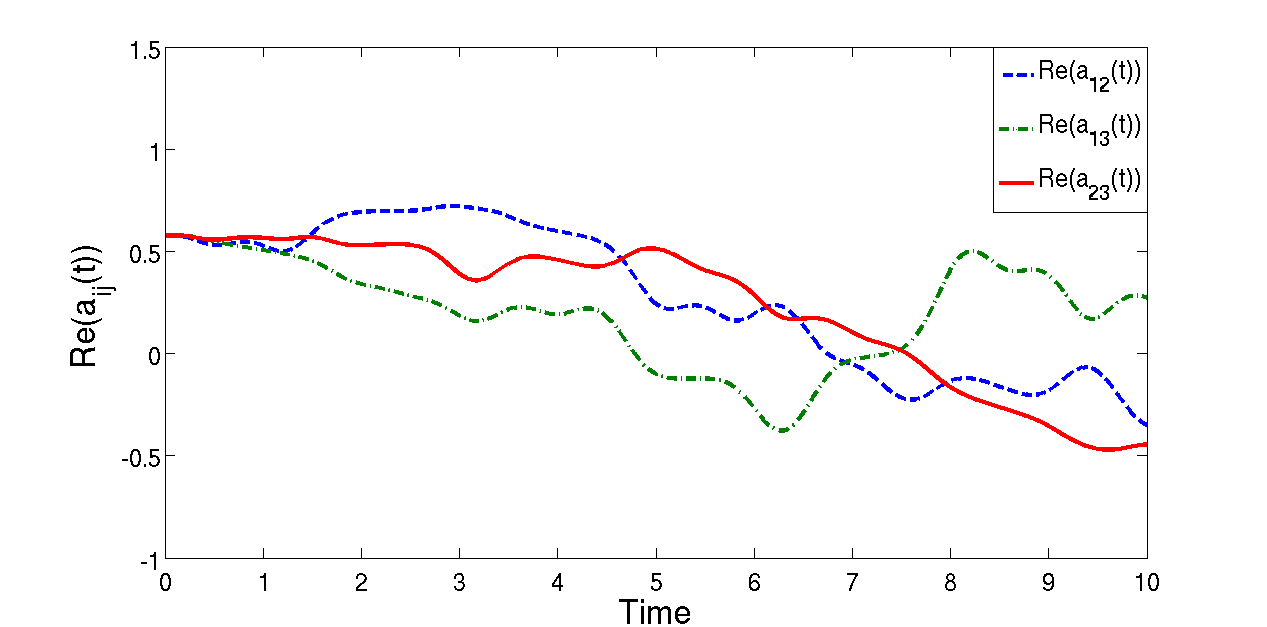} 
\includegraphics[width=1.0\columnwidth]{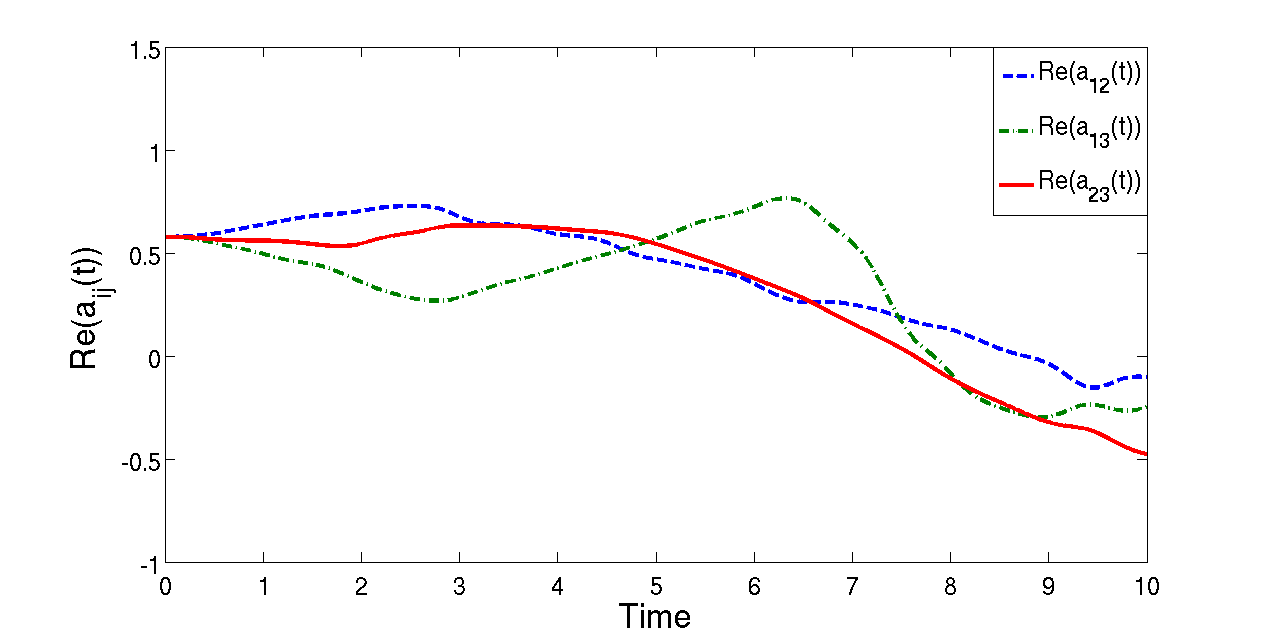}  \\ \hspace{-15cm} \text{b)}  \\ [-.05cm]
\includegraphics[width=1.0\columnwidth]{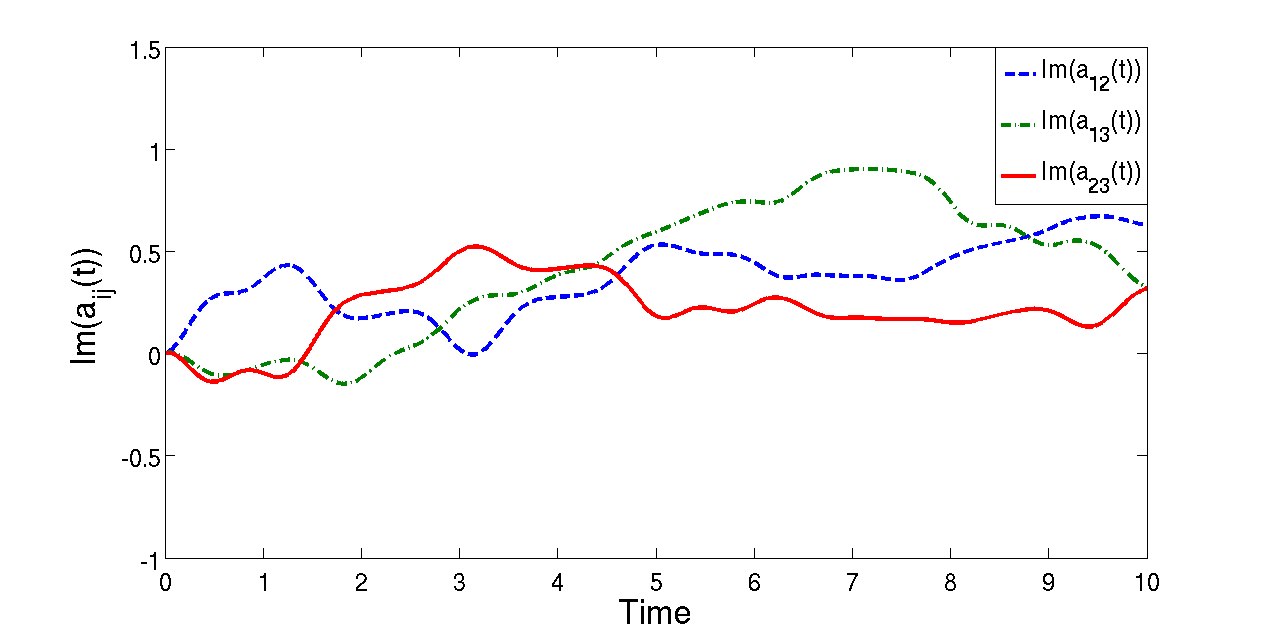} 
\includegraphics[width=1.0\columnwidth]{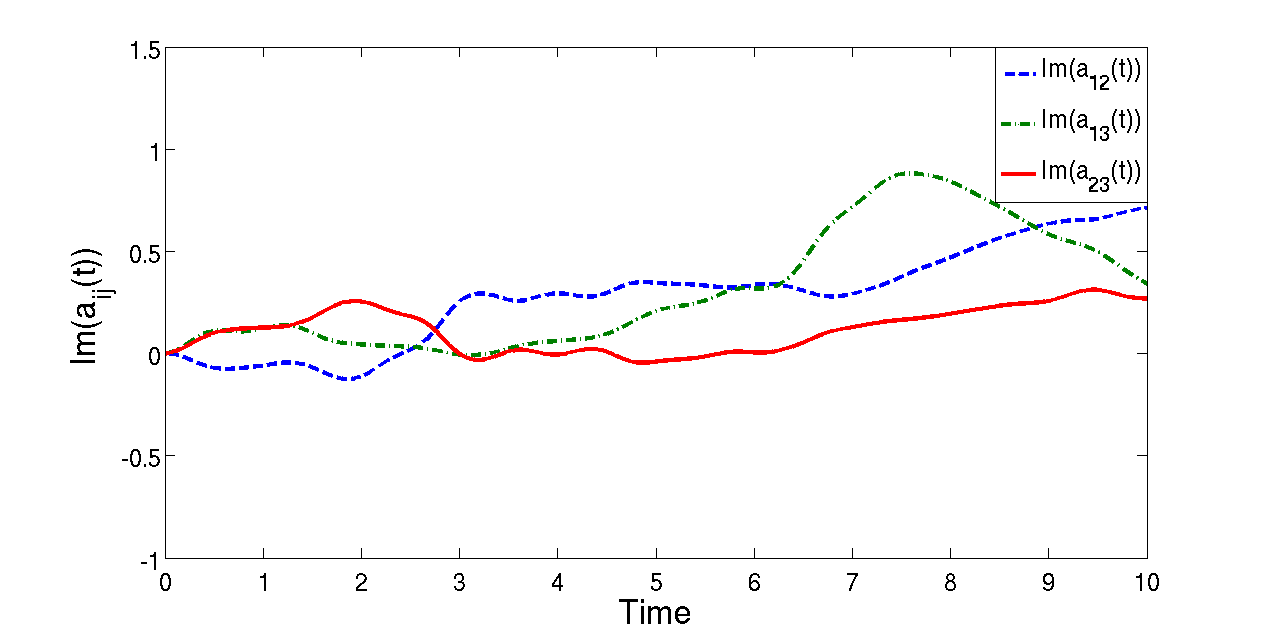}
\end{array}$
\end{centering}
\caption{\textbf{Wavefunctions in the Heisenberg Hamiltonian verses the XY Hamiltonian} - The real parts (a) and the imaginary parts (b) of the expansion coefficients of $|\psi(t)\rangle = \sum_{i<j }^3 a_{ij}(t) | i j \rangle $ (left column) and  $|\psi'(t)\rangle = \sum_{i<j }^3 a'_{ij}(t) | i j \rangle $ (right column) in the computational basis. Although the wavefunctions are clearly different, they both reproduce the same set $\{ \sigma_1^z, \sigma_2^z, \sigma_3^z \}$ throughout the evolution.}
\label{simulation_figure_3}
\end{figure*}

In this section we discuss how the proof of the VL theorem can be used to numerically construct a set of auxiliary fields $\{ h'_1, h'_2, ... h'_N \}$, which reproduce a given set $\{ \sigma_1^z, \sigma_2^z,... \sigma_N^z\}$ using a different two-qubit interaction. This procedure was used to simulate a Heisenberg model using an XY model in the main text, and was demonstrated in Figure~\ref{simulation_figure} and Figures~\ref{simulation_figure_2} and~\ref{simulation_figure_3} (below).

In principle, Eq.~\ref{recursion} can be used as a recursion relation to construct the Taylor coefficients of $\{ h'_1, h'_2, ... h'_N \}$ to arbitrary order, but in practice this proves to be numerically cumbersome. Instead, we use a formulation of the VL construction based on a non-linear Schr{\"o}dinger equation. A similar construction was presented in ref.~\cite{VL_extension} for electronic TDDFT.

We begin by numerically solving the time-dependent Schr{\"o}dinger equation in the "unprimed system",
\begin{equation}
\frac{\partial}{\partial t} |\psi(t) \rangle  = \hat{H}(t) |\psi(t) \rangle,
\label{original_TDSE}
\end{equation}
for a given initial state $|\psi(0)\rangle$ and a given Hamiltonian $\hat{H}(t)$ which we wish to simulate (see below for simulation details). From $|\psi(t) \rangle$, we can calculate all relevant observables, and in particular, we can calculate the set of currents $\{\langle \Hat{j}_{12} \rangle, \langle \Hat{j}_{13} \rangle, ... \langle \Hat{j}_{23} \rangle, ... \langle \Hat{j}_{N-1, N} \rangle \}$ at each time-step. 

We then construct the set of fields $\{ h'_1, h'_2, ... h'_N \}$ which reproduce this set of currents, but using a Hamiltonian with different two-qubit interactions $J'^{\perp}_{ij}$ and $J'^{\parallel}_{ij}$,
\begin{equation}
\hat{H}'(t) = \sum_{i < j}^N J'^{\perp}_{ij} (\hat{\sigma}_i^{x} \hat{\sigma}_j^{x} + \hat{\sigma}_i^{y} \hat{\sigma}_j^{y}) + \sum_{i < j}^N J'^{\parallel}_{ij} \hat{\sigma}_i^{z} \hat{\sigma}_j^{z} + \sum_{i=1}^N h'_i(t) \hat{\sigma}_i^z.
\label{primed qubit Hamiltonian}
\end{equation}
This is done by numerically solving Eq.~\ref{eom_current_primed} for the set $\{ h'_1, h'_2, ... h'_N \}$ at each time-step, with the requirement that $\{\langle \Hat{j}_{12} \rangle, \langle \Hat{j}_{13} \rangle, ... \langle \Hat{j}_{23} \rangle, ... \langle \Hat{j}_{N-1, N} \rangle \} = \{\langle \Hat{j}'_{12} \rangle', \langle \Hat{j}'_{13} \rangle', ... \langle \Hat{j}'_{23} \rangle', ... \langle \Hat{j}'_{N-1, N} \rangle' \}$ for all k and l. i.e. we solve
\begin{equation}
\frac{\partial}{\partial t} \langle \Hat{j}_{ki} \rangle = \langle \hat{\mathscr T'_{ki}} \rangle' + \langle \hat{\mathscr F'_{ki}} \rangle' + 4 \langle \hat{T}'_{ki} \rangle' \delta h'_{ik}(t),
\label{eom_current_primed_2}
\end{equation}
for all $\delta h'_{ik}(t)$, where on the left hand side $\langle \Hat{j}_{ki} \rangle$ are known from the solution to Eq.~\ref{original_TDSE}. However, since  $\langle \hat{\mathscr T'_{ki}} \rangle'$,  $\langle \hat{\mathscr F'_{ki}} \rangle'$ and $\langle \hat{T}'_{ki} \rangle' $ depend on the auxiliary wavefunction $|\psi'(t)\rangle$, to solve Eq.~\ref{eom_current_primed_2} we must simultaneously solve the auxiliary system's time-dependent Schr{\"o}dinger equation,
\begin{equation}
\frac{\partial}{\partial t} |\psi'(t) \rangle  = \hat{H}'(t) |\psi'(t) \rangle,
\label{NLSE}
\end{equation}
where $\hat{H}'(t)$ in turn depends on $\delta h'_{ik}(t)$. Thus, Eq.~\ref{NLSE} is a non-linear Schr{\"o}dinger equation and Eq.'s~\ref{eom_current_primed_2} and~\ref{NLSE} represent a set of coupled non-linear ordinary differential equations for the auxiliary wavefunction $|\psi'(t)\rangle$ and the set of fields $\{ h'_1, h'_2, ... h'_N \}$. To solve this system of equations, we begin with an initial state $|\psi'(0)\rangle$ which satisfies the conditions in Eq.'s~\ref{current_condition} and~\ref{density_condition}. By enforcing Eq.~\ref{density_condition}, we ensure that our solutions to Eq.'s~\ref{eom_current_primed_2} and~\ref{NLSE} will reproduce the set $\{ \sigma_1^z, \sigma_2^z,... \sigma_N^z\}$ in addition to the currents. From $|\psi'(0) \rangle$, we can solve Eq.~\ref{eom_current_primed_2} at $t=0$,
\begin{equation}
\frac{\partial}{\partial t} \langle \Hat{j}_{ki} \rangle|_{t=0} = \langle \hat{\mathscr T'_{ki}} \rangle'|_{t=0} + \langle \hat{\mathscr F'_{ki}} \rangle'|_{t=0} + 4 \langle \hat{T}'_{ki} \rangle'|_{t=0} \delta h'_{ik}(0),
\label{eom_current_primed_3}
\end{equation}
for field differences $\delta h'_{ik}(0)$. After making a choice for the arbitrary global field, we can construct the set of fields $\{ h'_1(0), h'_2(0), ...h'_N(0) \}$ at $t=0$ and the Hamiltonian $\hat{H}'(0)$. We then solve
\begin{equation}
\frac{\partial}{\partial t} |\psi'(t) \rangle |_{t=0} = \hat{H}'(0) |\psi'(0) \rangle,
\end{equation}
to obtain $|\psi'(dt)\rangle$ at the next time-step. From $|\psi'(dt)\rangle$, we obtain $\{ h'_1(dt), h'_2(dt), ...h'_N(dt) \}$ by solving
\begin{equation}
\frac{\partial}{\partial t} \langle \Hat{j}_{ki} \rangle|_{t=dt} = \langle \hat{\mathscr T'_{ki}} \rangle'|_{t=dt} + \langle \hat{\mathscr F'_{ki}} \rangle'|_{t=dt} + 4 \langle \hat{T}'_{ki} \rangle'|_{t=dt} \delta h'_{ik}(dt).
\label{eom_current_primed_4}
\end{equation}
This procedure is continued at each time-step until we obtain $\{ h'_1, h'_2, ...h'_N \}$ and $|\psi'(t) \rangle$ on the entire interval [0,t].


For the simulation presented in the main text, $\hat{H}(t)$ is the Heisenberg Hamiltonian and $\hat{H}'(t)$ is an XY Hamiltonian with the chosen parameters. With the initial state $|\psi(0)\rangle = \frac{1}{\sqrt{3}}(|011\rangle + |101\rangle + |110\rangle)$, we solve Eq.~\ref{original_TDSE} using the fourth-order Runge-Kutta method. $|\psi(t) \rangle$ is propagated on a uniform grid with $10^{4}$ time-steps, each of duration $dt = 1.5 \times 10^{-4} \frac{\hbar}{2J}$. With $|\psi(t)\rangle$, we can calculate derivatives of the currents between all 3 qubits to be used in Eq.~\ref{eom_current_primed_2}. We use the procedure outlined in Eq.'s~\ref{eom_current_primed_2}-~\ref{eom_current_primed_4} to obtain $|\psi'(t)\rangle$ and $\{ h'_1, h'_2, h'_3 \}$ of the auxiliary XY Hamiltonian. For the auxiliary system's initial state, we chose $|\psi'(0) \rangle = |\psi(0)\rangle = \frac{1}{\sqrt{3}}(|011\rangle + |101\rangle + |110\rangle)$, which satisfies the conditions in Eq.'s~\ref{current_condition} and~\ref{density_condition} since the initial currents vanish in both the primed and unprimed systems. We also fix the arbitrary global field by choosing $h'_2(t) = h_2(t)$ for all t. Since $|\psi'(0)\rangle$ is an eigenstate of $\hat{\sigma}^z_{total}$, this choice corresponds to trivially fixing the global phase of the auxiliary system's wavefunction and any other choice would yield identical expectation values of observables. 

As shown in Figure~\ref{simulation_figure}, the set $\{ \sigma_1^z, \sigma_2^z, \sigma_3^z\}$ is faithfully reproduced by $|\psi'(t) \rangle$. In Figure~\ref{simulation_figure_2} we show expectation values of several other observables calculated with $|\psi(t) \rangle$ in the left column and $|\psi'(t) \rangle$ in the right column. Naturally, observables that depend explicitly on the set $\{ \sigma_1^z, \sigma_2^z,\sigma_3^z\}$ are the same in both cases, while those that do not will be different. In particular, we see that the entanglement is the same in both cases, since both $|\psi(t) \rangle$ and $|\psi'(t) \rangle$ remain a superposition of states with one flipped qubit during the evolution. This means that the explicit entanglement functional in Eq.~\ref{entanglement} holds, and since both wavefunctions produce the same set $\{ \sigma_1^z, \sigma_2^z,\sigma_3^z\}$, they necessarily produce the same entanglement. As expected, the currents $\{\langle \Hat{j}_{12} \rangle, \langle \Hat{j}_{23} \rangle \}$ and $\{\langle \Hat{j}'_{12} \rangle', \langle \Hat{j}'_{23} \rangle' \}$ are the same for both wavefunctions, while the kinetic terms $\{\langle \hat{T}_{12} \rangle, \langle \hat{T}_{23} \rangle \}$ and $\{\langle \hat{T}'_{12} \rangle', \langle \Hat{T}'_{23} \rangle' \}$ are different. The same situation arises in electronic DFT, where the Kohn-Sham wavefunction reproduces the correct density and current, but the kinetic energy is in general different from that of the true correlated wavefunction. In Figure~\ref{simulation_figure_3}, we plot the expansion coefficients of $|\psi(t) \rangle$ and $|\psi'(t) \rangle$ in the computational basis $\{ |011\rangle, |101\rangle, |110\rangle \}$, which as expected are rather different. It is also interesting to note that in our formalism, the operators for the current and kinetic energy are different in the original and auxiliary systems, since we let all two-qubit parameters in the Hamiltonian differ. This situation is different than in electronic TDDFT, where the kinetic energy and current operators themselves are the same, although expectation values may be different in the case of the kinetic energy. This is important in our formalism, especially with regard to the current, since although $\{\langle \Hat{j}_{12} \rangle, \langle \Hat{j}_{23} \rangle \} = \{\langle \Hat{j}'_{12} \rangle', \langle \Hat{j}'_{23} \rangle' \}$, one finds that in general $\{\langle \Hat{j}_{12} \rangle, \langle \Hat{j}_{23} \rangle \} \neq \{\langle \Hat{j}_{12} \rangle', \langle \Hat{j}_{23} \rangle' \}$.

\begin{figure*}[h]
\begin{centering}
$\begin{array}{c@{\hspace{1in}}c}
 \\[.2 cm] \hspace{1.6cm} \text{\underline{\Large{Electronic TDDFT}} \hspace{3.0cm} \underline{\Large{TDDFT for Quantum Computation}}} \\ 
\includegraphics[width=1.0\columnwidth]{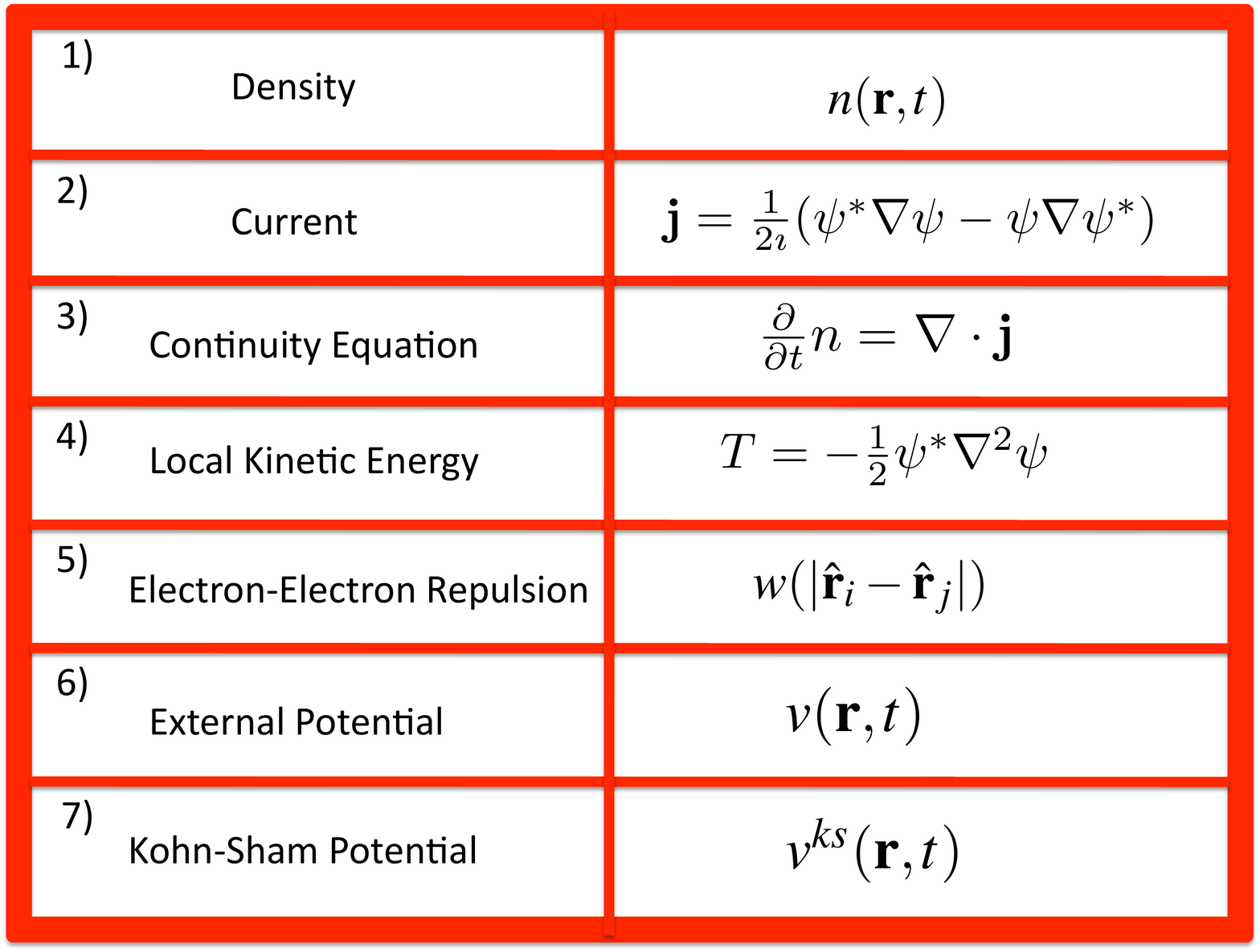} 
\includegraphics[width=1.0\columnwidth]{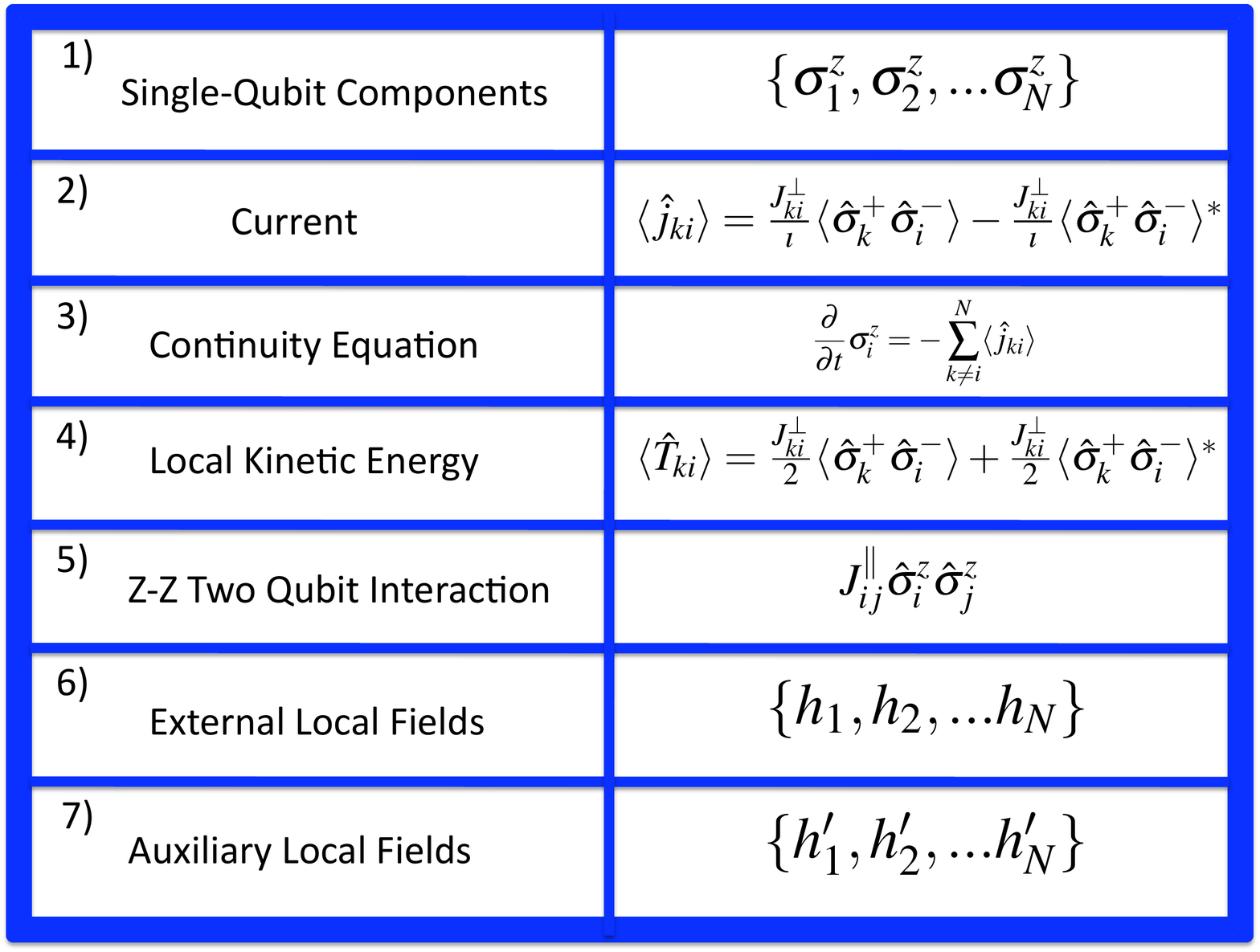}   \end{array}$
\end{centering}
\caption{\textbf{Analogies between electronic TDDFT and TDDFT for quantum computation} - Relevant quantities in electronic TDDFT (left table) and the corresponding quantities in TDDFT for quantum computation (right table).}
\label{comparison tables}
\end{figure*}

\subsection{Analogies between electronic TDDFT and TDDFT for quantum computation.}

Throughout the manuscript we have tried to stress the analogies between electronic TDDFT and TDDFT for systems of qubits. The main relevant quantities in electronic TDDFT and the analogous quantities in TDDFT for quantum computation are summarized in Figure~\ref{comparison tables}. The current and local kinetic energy for electronic TDDFT are written for a single electron, but the extension to N electrons is straightforward, by integrating over $N-1$ coordinates.

Despite the clear similarities, there are also important differences between qubit and electronic systems. Qubits are distinguishable quantum particles and the wavefunction does not need to obey any particular permutational symmetry. In contrast, electrons are indistinguishable fermions with a fully antisymmetric wavefunction. Through the Jordan-Wigner transformation~\cite{Jordan_Wigner}, a system of qubits can in fact be mapped into a system of spinless fermions, but we do not pursue this in the present work. Also, the pauli sigma operators obey a different commutator algebra than the electronic position and momentum operators. The resulting kinetic energy operator is a two-qubit quantity in TDDFT for quantum computation, while for electronic TDDFT it is a one-electron operator.




\end{document}